\documentclass[letter,a4paper,twocolumn]{jpsj3}
\usepackage{txfonts}
\usepackage{bm}
\usepackage{color}
\usepackage{mathrsfs}

\title{Raman Scattering Polarization and Single Spinon Identification in\\
       Two-Dimensional Kitaev Quantum Spin Liquids}
\author{Shoji Yamamoto${}^*$ and Taku Kimura}
\inst{Department of Physics, Hokkaido University, Sapporo 060-0810, Japan} 
\abst{Perfect deporalization of the Loudon-Fleury inelastic visible-light scattering in
      the Kitaev honeycomb model is well known.
      Though it happens in Heisenberg Kagome and triangular antiferromagnets as well,
      yet we prove it to be of geometric origin rather than peculiar to quantum spin liquids.
      A Kitaev spin liquid in the square planar geometry indeed exhibits polarized Raman spectra
      containing defferent symmetry species, each brought by symmetry-compatible spinon geminate
      excitations, i.e. arising from symmetry-compatible direct-product representations made of
      double-valued irreducible representations of mediating-spinon-belonging
      gauged $\bm{k}$-point symmetry groups.
      We combine a standard point-symmetry-group analysis of the Raman vertex in the real space
      and an elaborate projective-symmetry-group analysis of Raman-scattering-mediating Majorana
      spinons in the reciprocal space to identify emergent spinons singly.
      \vspace*{-2mm}}

\begin{document}
\maketitle

   The Kitaev honeycomb model \cite{K2} has made a major breakthrough in the study of quantum spin
liquids (QSLs), \cite{K451,M012002} explicitly visualizing partons \cite{S016502} as elementary
excitations.
Majorana spinons accompanied by emergent $\mathbb{Z}_2$ gauge fields are characteristic of
the Kitaev QSL and Raman spectroscopy is particularly useful in diagnosing them. \cite{K187201}
Within the Loudon-Fleury (LF) mechanism \cite{F514} valid for strongly correlated electrons,
\cite{S1068,S365} the Raman vertex commutes with background gauge fields and can selectively
excite spinons. \cite{K187201,P094439}

   The LF vertex for the gauge-ground Kitaev honeycomb QSL yields a completely depolarized Raman
response, \cite{K187201,Y012003} but the depolarization is no longer perfect beyond the LF theory,
\cite{P060408,P104427}
in an external field, \cite{P060408,P104427,P184429}
and with integrability-breaking perturbations such as Heisenberg \cite{K187201,R045117}
and off-diagonal \cite{R045117} exchanges,
whether intralayer or interlayer. \cite{R077204,T094403,S115159}
On the other hand, the depolarization of the LF Raman response occurs in
Heisenberg frustrated antiferromagnets as well, \cite{C172406,P174412}
including a $\mathrm{U}(1)$ Dirac spin-liquid state on the regular Kagome lattice \cite{K024414}
and a $\mathbb{Z}_2$ dimer-liquid phase on the equilateral triangular lattice \cite{M134421}.
There is an argument that depolarization of Raman response may be characteristic of QSLs.
\cite{C172406}

\begin{figure}[b]
\vspace*{-4mm}
\centering
\includegraphics[width=86mm]{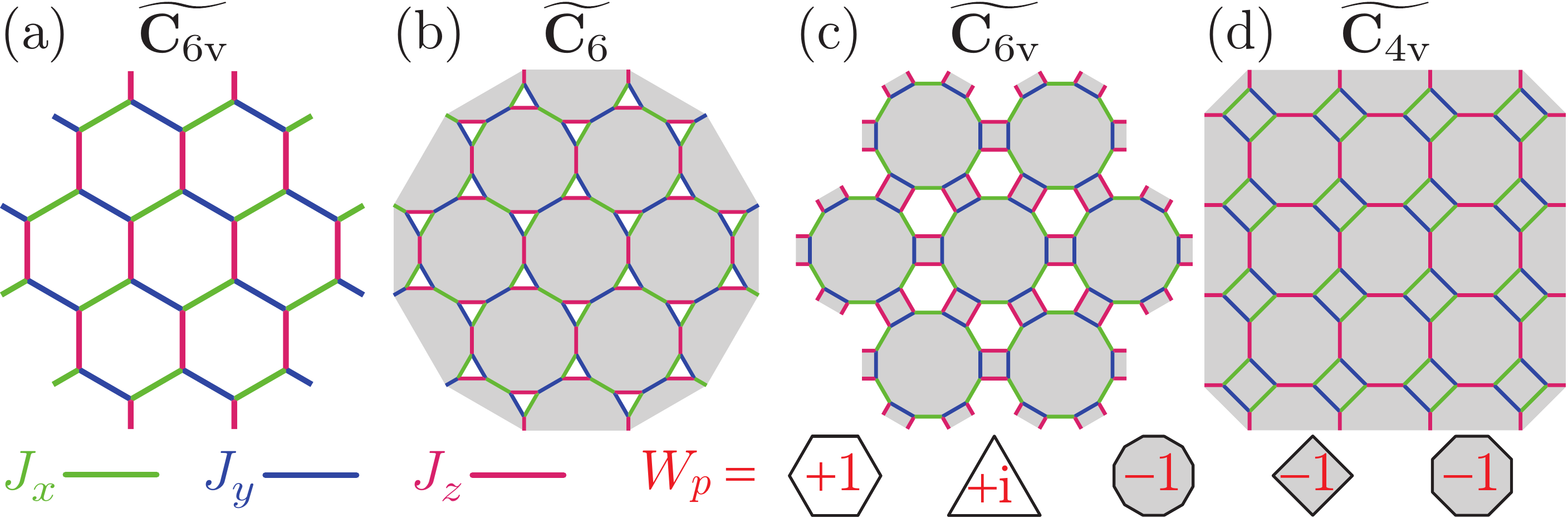}
\vspace*{-5mm}
\caption{(Color online)
         Kitaev models consisting of the pure (a)-, triangle (b)-, square-hexagon (c)-honeycomb
         and diamond-square (d) lattices in their ground flux configurations.
         In (b), the constituent triangles are arrangeable into
         either $\{W_p=+i;\,p=1,\cdots,\frac{L}{3}\}$
         or $\{W_p=-i;\,p=1,\cdots,\frac{L}{3}\}$.}
\label{F:2DKitaevModels}
\end{figure}

   We are thus motivated to discuss Raman responses of various two-dimensional Kitaev models
(Fig. \ref{F:2DKitaevModels}).
We make symmetry arguments in two ways.
Point-symmetry \cite{SM:prsym} analysis of the LF vertex [cf. Eq. (\ref{E:I(w)LF})]
in the real space reveals what is the decisive factor for Raman scattering
polarization, and then, projective-symmetry \cite{W165113,W174423,SM:prsym} analysis of
the gauge-ground Majorana Hamiltonian [cf. Eq. (\ref{E:HMajoranaGSr})] in the reciprocal space
shows which mode of polarized Raman responses, if any, is attributable to which combination
of spinon eigenmodes.

   The Kitaev Hamiltonian (cf. Fig. \ref{F:2DKitaevModels}) reads
\begin{align}
   \mathscr{H}
 =-\sum_{<\bm{r}_l:\lambda,\bm{r}_{l'}:\lambda'>}
   J_{\alpha(\bm{r}_l:\lambda,\bm{r}_{l'}:\lambda')}
   \sigma_{\bm{r}_l:\lambda}^{\alpha(\bm{r}_l:\lambda,\bm{r}_{l'}:\lambda')}
   \sigma_{\bm{r}_{l'}:\lambda'}^{\alpha(\bm{r}_l:\lambda,\bm{r}_{l'}:\lambda')}
   \label{E:Hspin}
   \\[-8mm]\nonumber
\end{align}
where
$(\sigma_{\bm{r}_l:\lambda}^x,\sigma_{\bm{r}_l:\lambda}^y,\sigma_{\bm{r}_l:\lambda}^z)$
$(l=1,\cdots,N^2\equiv L/q;\,\lambda=1,\cdots,q)$ are the Pauli matrices attached to
the $\lambda$th site in the $l$th unit at $\bm{r}_l$ and obey the usual commutation relations
$[\sigma_{\bm{r}_l:\lambda}^\beta,\sigma_{\bm{r}_{l'}:\lambda'}^\gamma]
=2i\delta_{ll'}\delta_{\lambda\lambda'}
 \sum_{\alpha=x,y,z}\epsilon_{\alpha\beta\gamma}\sigma_{\bm{r}_l:\lambda}^\alpha$,
while
$<\bm{r}_l:\lambda,\bm{r}_{l'}:\lambda'>$ runs over $3L/2$ nearest-neighbor bonds with
$\alpha(\bm{r}_l:\lambda,\bm{r}_{l'}:\lambda')$ as a function of $\bm{r}_l:\lambda$ and
$\bm{r}_{l'}:\lambda'$
taking $x$, $y$, and $z$ once every $\bm{r}_l:\lambda$.
The coupling constants $J_{\alpha(\bm{r}_l:\lambda,\bm{r}_{l'}:\lambda')}$
are all set to $J>0$ in the following.
We start with the pure honeycomb lattice containing $L/2$ hexagons
[Fig. \ref{F:2DKitaevModels}(a)],
next decorate this putting triangles on its vertices to have $L/3$ triangles and $L/6$
dodecahedrons [Fig. \ref{F:2DKitaevModels}(b)]
\cite{Y247203,N087203,Y180404} or hexagons and squares on its vertices and sides
to have $L/4$ squares, $L/6$ hexagons, and $L/12$ dodecahedrons
[Fig. \ref{F:2DKitaevModels}(c)], \cite{Y180404} and then proceed to a square analog
decorated with diamonds containing $L/4$ diamonds and $L/4$ octagons
[Fig. \ref{F:2DKitaevModels}(d)]. \cite{Y180404,K57007,B6918}
Denoting the primitive translation vectors in each lattice by $\bm{a}$ and $\bm{b}$,
we adopt a periodic boundary condition, $\bm{r}_l+N\bm{a}=\bm{r}_l+N\bm{b}=\bm{r}_l$.
The number of sites $L$ reads $qN^2$ with
$q=2$ [Fig. \ref{F:2DKitaevModels}(a)],
$q=6$ [Fig. \ref{F:2DKitaevModels}(b)],
$q=12$ [Fig. \ref{F:2DKitaevModels}(c)], and
$q=4$ [Fig. \ref{F:2DKitaevModels}(d)].

   We introduce four Majorana fermions at each site as
$\sigma_{\bm{r}_l:\lambda}^\alpha=i\eta_{\bm{r}_l:\lambda}^\alpha c_{\bm{r}_l:\lambda}$ with
$\{\eta_{\bm{r}_l:\lambda}^\beta,\eta_{\bm{r}_{l'}:\lambda'}^\gamma\}
=2\delta_{ll'}\delta_{\lambda\lambda'}\delta_{\beta\gamma}$,
$\{c_{\bm{r}_l:\lambda},c_{\bm{r}_{l'}:\lambda'}\}=2\delta_{ll'}\delta_{\lambda\lambda'}$,
and
$\{\eta_{\bm{r}_l:\lambda}^\alpha,c_{\bm{r}_{l'}:\lambda'}\}=0$
to have
\begin{align}
   \mathscr{H}
  =iJ\sum_{<\bm{r}_l:\lambda,\bm{r}_{l'}:\lambda'>}
   \hat{u}_{<\lambda,\lambda'>}
          ^{\bm{r}_l\to\bm{r}_{l'}}
   c_{\bm{r}_l:\lambda}c_{\bm{r}_{l'}:\lambda'},
   \label{E:HMajorana}
\end{align}
where the nearest-neighbor bond operators
$
   \hat{u}_{<\lambda,\lambda'>}^{\bm{r}_l\to\bm{r}_{l'}}
  \equiv
  i\eta_{\bm{r}_l:\lambda}^{\alpha(\bm{r}_l:\lambda,\bm{r}_{l'}:\lambda')}
   \eta_{\bm{r}_{l'}:\lambda'}^{\alpha(\bm{r}_l:\lambda,\bm{r}_{l'}:\lambda')}
 =-\hat{u}_{<\lambda',\lambda>}^{\bm{r}_{l'}\to\bm{r}_l}
$
commute with each other as well as the Hamiltonian (\ref{E:HMajorana}) and therefore
behave as $\mathbb{Z}_2$ classical variables,
$
   u_{<\lambda,\lambda'>}^{\bm{r}_l\to\bm{r}_{l'}}
  =\pm 1
$.
For an $N_p$-sided polygon, we multiply its constituent spin operators in the anticlockwise
manner to define the flux operator \cite{K2,H035145,P134404,Z014403}
\begin{align}
   \hat{W}_p
   &
  \equiv
   \prod_{<\bm{r}_l:\lambda,\bm{r}_{l'}:\lambda'>\in\partial p}
   \sigma_{\bm{r}_l:\lambda}^{\alpha(\bm{r}_l:\lambda,\bm{r}_{l'}:\lambda')}
   \sigma_{\bm{r}_{l'}:\lambda'}^{\alpha(\bm{r}_l:\lambda,\bm{r}_{l'}:\lambda')}
   \nonumber \\
   &
  =(-i)^{N_p}
   \prod_{<\bm{r}_l:\lambda,\bm{r}_{l'}:\lambda'>\in\partial p}
   \hat{u}_{<\lambda,\lambda'>}^{\bm{r}_l\to\bm{r}_{l'}}
   \label{E:Wp}
\end{align}
$\hat{W}_p$ also commutes with the Hamiltonian,
whether (\ref{E:Hspin}) or (\ref{E:HMajorana}), and thus behaves as a classical variable,
$W_p=\pm 1$ or $\pm i$ according as $N_p$ is even or odd.
We have a $\mathrm{U(1)}$ gauge flux,
$W_p\equiv e^{i\varPhi_p}\,(-\pi<\varPhi_p\leq\pi)$,
in general.
Each Kitaev lattice consists of $\frac{L}{2}$ gauged polygons with their flux
variables satisfying $\prod_{p=1}^{L/2}W_p=1$. \cite{K2}
Under the periodic boundary condition, there are two more nontrivial flux operators
$\hat{W}_c\,(c=a,b)$ \cite{H035145,Z014403} wrapping around the torus in the directions
$\bm{a}$ and $\bm{b}$, each with eigenvalues
$\pm 1$ in Figs. \ref{F:2DKitaevModels}(a)--\ref{F:2DKitaevModels}(c) and
$\pm 1$ or $\pm i$ according as $N$ is even or odd in Fig. \ref{F:2DKitaevModels}(d).
We set $N$ equal to a sufficiently large even number, $8192$ or more.
We have $2^{\frac{L}{2}+1}$ flux configurations $\{W_p,W_c\}$,
each available from a set of $2^{\frac{3L}{2}}/2^{\frac{L}{2}+1}=2^{L-1}$ different
bond configurations
$\{u_{<\lambda,\lambda'>}^{\bm{r}_l\to\bm{r}_{l'}}\}$.
The eigenspectrum of (\ref{E:HMajorana}) depends on
$\{u_{<\lambda,\lambda'>}^{\bm{r}_l\to\bm{r}_{l'}}\}$ only through $\{W_p,W_c\}$.

   The ground-flux-configuration sector of (\ref{E:HMajorana}) reads
\begin{align}
   \mathscr{H}
  =\frac{iJ}{2}
   \sum_{l=1}^{N^2}
   \sum_{\bm{\delta}=\bm{0},\pm\bm{a},\pm\bm{b}}
   \sum_{\lambda=1}^q\sum_{\lambda'=1}^q
   u_{<\lambda,\lambda'>}^{\bm{\delta}}
   c_{\bm{r}_l:\lambda}c_{\bm{r}_l+\bm{\delta}:\lambda'},
   \label{E:HMajoranaGSr}
\end{align}
where
$
   u_{<\lambda,\lambda'>}^{\bm{\delta}}
  \equiv
   u_{<\lambda,\lambda'>}^{\bm{r}_l\to\bm{r}_l+\bm{\delta}}
$
no longer depend on the position $\bm{r}_l$.
Carrying out the Fourier transformation
\begin{align}
   \gamma_{\bm{k}_\kappa:\lambda}
  =\frac{1}{\sqrt{2}N}\sum_{l=1}^{N^2}
   e^{i\bm{k}_\kappa\cdot\bm{r}_l}
   c_{\bm{r}_l:\lambda},\ 
   c_{\bm{r}_l:\lambda}
  =\frac{\sqrt{2}}{N}\sum_{\kappa=1}^{N^2}
   e^{-i\bm{k}_\kappa\cdot\bm{r}_l}
   \gamma_{\bm{k}_\kappa:\lambda}
   \label{E:FT}
\end{align}
with $\gamma_{-\bm{k}_\kappa:\lambda}=\gamma_{\bm{k}_\kappa:\lambda}^\dagger$ in mind yields
\begin{align}
   &
   \mathscr{H}
  =iJ\sum_{\kappa=1}^{N^2}
   \sum_{\lambda=1}^q\sum_{\lambda'=1}^q
   u_{<\lambda,\lambda'>}(\bm{k}_\kappa)
   \gamma_{\bm{k}_\kappa:\lambda}^\dagger\gamma_{\bm{k}_\kappa:\lambda'};
   \nonumber
   \allowdisplaybreaks
   \\
   &
   u_{<\lambda,\lambda'>}(\bm{k}_\kappa)
  \equiv
   \sum_{\bm{\delta}=\bm{0},\pm\bm{a},\pm\bm{b}}
   u_{<\lambda,\lambda'>}^{\bm{\delta}}
   e^{-i\bm{k}_\kappa\cdot\bm{\delta}}
  =u_{<\lambda,\lambda'>}^*(-\bm{k}_\kappa).
   \label{E:HMajoranaGSk}
\end{align}
We define $q/2$ (complex) bond fermions
\begin{align}
   f_{ \bm{k}_\kappa:\lambda}
  \equiv
   \frac{\gamma_{\bm{k}_\kappa:2\lambda-1}+i\gamma_{\bm{k}_\kappa:2\lambda}}{\sqrt{2}},\ 
   f_{-\bm{k}_\kappa:\lambda}^\dagger
  =\frac{\gamma_{\bm{k}_\kappa:2\lambda-1}-i\gamma_{\bm{k}_\kappa:2\lambda}}{\sqrt{2}}
\end{align}
at each momentum to process (\ref{E:HMajoranaGSk}) into
\vspace*{-1mm}
\begin{align}
   \mathscr{H}
  =\sum_{\kappa=1}^{N^2}
   \mathcal{F}_{\bm{k}_\kappa}^\dagger
   \mathcal{H}_{\bm{k}_\kappa}
   \mathcal{F}_{\bm{k}_\kappa}
  \equiv
   \sum_{\kappa=1}^{N^2}
   \mathcal{F}_{\bm{k}_\kappa}^\dagger
   \left[
    \begin{array}{cc}
     \!\!\!  \mathcal{H}_{ \bm{k}_\kappa}^{(+)}       \!\!\! &
     \!\!\!  \mathcal{H}_{ \bm{k}_\kappa}^{(-)}       \!\!\! \\
     \!\!\!  \mathcal{H}_{ \bm{k}_\kappa}^{(-)\dagger}\!\!\! &
     \!\!\! -\mathcal{H}_{-\bm{k}_\kappa}^{(+)*}      \!\!\! \\
    \end{array}
   \right]
   \mathcal{F}_{\bm{k}_\kappa}
   \label{E:HMajoranaGSkF}
\end{align}
with vectors of dimension $q$ and matrices of dimension $\frac{q}{2}\times\frac{q}{2}$
\begin{align}
   &
   \mathcal{F}_{\bm{k}_\kappa}
  \equiv
   \Bigl[
    f_{ \bm{k}_\kappa:1}^\dagger\cdots f_{ \bm{k}_\kappa:\frac{q}{2}}^\dagger
    f_{-\bm{k}_\kappa:1}        \cdots f_{-\bm{k}_\kappa:\frac{q}{2}}
   \Bigr]^\dagger,
   \allowdisplaybreaks
   \\
   &
   \left[
    \mathcal{H}_{\bm{k}_\kappa}^{(\sigma)}
   \right]_{\lambda\lambda'}
  \equiv
   \frac{iJ}{2}
   \left[
            u_{<2\lambda-1,2\lambda'-1>}(\bm{k}_\kappa)
   -i\sigma u_{<2\lambda-1,2\lambda'  >}(\bm{k}_\kappa)
   \right.
   \nonumber \\
   &\qquad\qquad\qquad
   \left.
   +i       u_{<2\lambda  ,2\lambda'-1>}(\bm{k}_\kappa)
   + \sigma u_{<2\lambda  ,2\lambda'  >}(\bm{k}_\kappa)
   \right],
   \nonumber
   \allowdisplaybreaks
   \\
   &
   \left[
    \mathcal{H}_{\bm{k}_\kappa}^{(+)}
   \right]_{\lambda\lambda'}
  =\left[
    \mathcal{H}_{\bm{k}_\kappa}^{(+)}
   \right]_{\lambda'\lambda}^*,\ 
   \left[
    \mathcal{H}_{ \bm{k}_\kappa}^{(-)}
   \right]_{\lambda\lambda'}
 =-\left[
    \mathcal{H}_{-\bm{k}_\kappa}^{(-)}
   \right]_{\lambda'\lambda}.
\end{align}
Having in mind that the $\pm\bm{k}_\kappa$ blocks $\mathcal{H}_{\pm\bm{k}_\kappa}$
of the Hamiltonian (\ref{E:HMajoranaGSkF}) are related through two different unitary
transformations,
\begin{align}
   \mathcal{H}_{\bm{k}_\kappa}
  =-\mathcal{M}^\dagger \ \!^{\mathrm{t}}\mathcal{H}_{-\bm{k}_\kappa} \mathcal{M}
  =\widetilde{C_2}^\dagger \mathcal{H}_{-\bm{k}_\kappa} \widetilde{C_2},
\end{align}
where $\mathcal{M}$ reads
$[\mathcal{M}]_{\lambda,\lambda+\frac{q}{2}}=[\mathcal{M}]_{\lambda+\frac{q}{2},\lambda}=1$
$(\lambda=1,\cdots,\frac{q}{2})$ and otherwise consists of $0$,
while
$\widetilde{C_2}$ is a gauged twofold rotation, \cite{SM:prsym}
we find that $\mathcal{H}_{\pm\bm{k}_\kappa}$ have the same set of eigenvalues and
every such set consists of $q/2$ pairs of eigenvalues
$\pm\epsilon_{\bm{k}_\kappa:\lambda}$ $(\lambda=1,\cdots,\frac{q}{2})$. \cite{O085101}
Arranging creation operators for spinon particles and holes into a column vector,
\vspace*{-1mm}
\begin{align}
   \mathcal{A}_{\bm{k}_\kappa}
  \equiv
   \Bigl[
    \alpha_{\bm{k}_\kappa:1}^\dagger\cdots\alpha_{\bm{k}_\kappa:\frac{q}{2}}^\dagger
    \alpha_{\bm{k}_\kappa:1}        \cdots\alpha_{\bm{k}_\kappa:\frac{q}{2}}
   \Bigr]^\dagger,
\end{align}
and defining their energies as
\begin{align}
   \mathcal{E}_{\bm{k}_\kappa}
  \equiv
   \mathrm{diag}
   \bigl[
    \varepsilon_{\bm{k}_\kappa:1},\cdots, \varepsilon_{\bm{k}_\kappa:\frac{q}{2}},
   -\varepsilon_{\bm{k}_\kappa:1},\cdots,-\varepsilon_{\bm{k}_\kappa:\frac{q}{2}}   
   \bigr],
\end{align}
the gauge-ground Hamiltonian (\ref{E:HMajoranaGSr}) reads
\vspace*{-1mm}
\begin{align}
   \mathscr{H}
  =\sum_{\kappa=1}^{N^2}
   \mathcal{A}_{\bm{k}_\kappa}^\dagger
   \frac{\mathcal{E}_{\bm{k}_\kappa}}{2}
   \mathcal{A}_{\bm{k}_\kappa}
  \equiv
   \sum_{\kappa=1}^{N^2}\sum_{\lambda=1}^{q/2}
   \varepsilon_{\bm{k}_\kappa:\lambda}
   \left(
    \alpha_{\bm{k}_\kappa:\lambda}^\dagger\alpha_{\bm{k}_\kappa:\lambda}-\frac{1}{2}
   \right),
   \label{E:HMajoranaGSkA}
\end{align}
where the eigenvalues
$\varepsilon_{\bm{k}_\kappa:\lambda}\equiv|\epsilon_{\bm{k}_\kappa:\lambda}|$ are nonnegative.

\begin{figure*}[t]
\centering
\includegraphics[width=177mm]{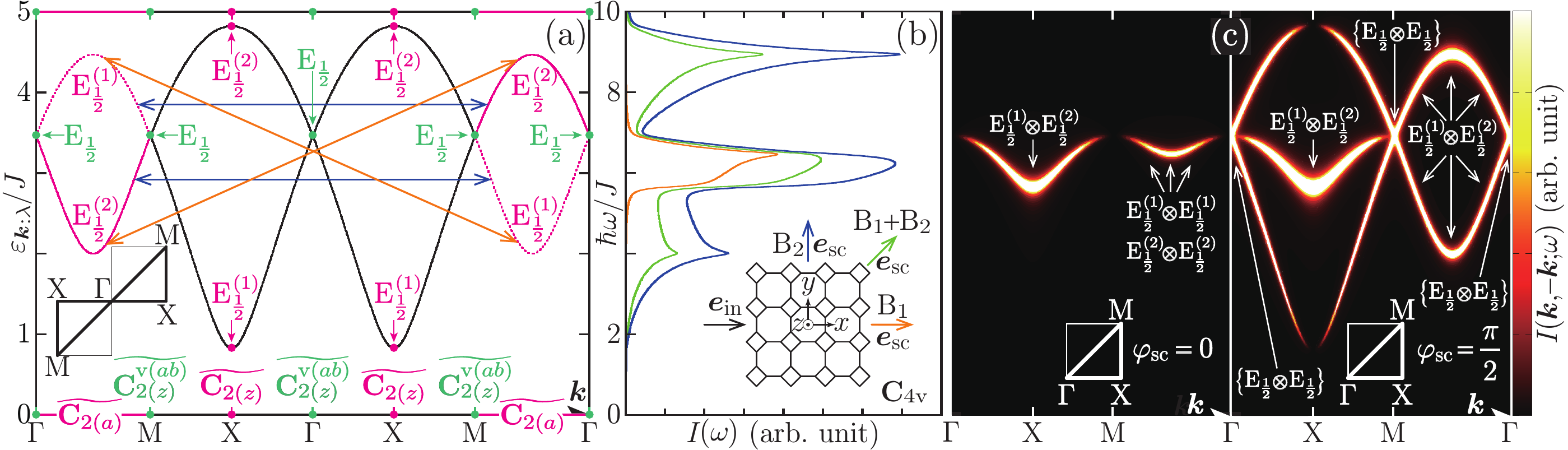}
\vspace*{-5mm}
\caption{(Color online)
         Spinon excitation energies $\varepsilon_{\bm{k}:\lambda}$ $(\lambda=1,2)$ (a)
         and Raman intensities
         $I(\omega)\equiv\sum_{\kappa=1}^{N^2}I(\bm{k}_\kappa,-\bm{k}_\kappa;\omega)$
         with $\varphi_{\mathrm{in}}=0$ and
         $\varphi_{\mathrm{sc}}=0,\frac{\pi}{4},\frac{\pi}{2}$ (b) of the gauge-ground Kitaev
         diamond-square model.
         The momentum-resolved spectral weights $I(\bm{k},-\bm{k};\omega)$ (c) arise from
         the spinon geminate excitations
         $\sum_{\lambda,\lambda'=1}^2
          \alpha_{\bm{k}:\lambda}^\dagger\alpha_{-\bm{k}:\lambda'}^\dagger$, each belonging to
         a direct-product representation of the gauged $\bm{k}$-point symmetry group
         $\widetilde{\mathbf{P}_{\bm{k}}'}$ [see (a) and Table \ref{T:DPRintoIrrep}].}
\label{F:RamanSquare}
\vspace*{-5mm}
\end{figure*}

   Given a lattice of point symmetry $\mathbf{P}_{\mathrm{org}}$, a wavevector $\bm{k}$ in
the first Brillouin zone of its reciprocal lattice belongs to the $\bm{k}$-point symmetry
group (isotropy group of $\bm{k}$) $\mathbf{P}_{\bm{k}}\subseteq\mathbf{P}_{\mathrm{org}}$,
\cite{Y044709} which consists of point symmetry operations
$P_{\bm{k}}$ such that $P_{\bm{k}}\bm{k}=\bm{k}+\bm{K}\cong\bm{k}$
with $\bm{K}$ being $\bm{0}$ or a reciprocal lattice vector.
Each $\mathcal{H}_{\bm{k}_\kappa}$ of the gauge-ground Majorana Hamiltonian
(\ref{E:HMajoranaGSkF}) may belong to a projective $\bm{k}$-point symmetry group
$\widetilde{\mathbf{P}_{\bm{k}}'}$, \cite{SM:Irreps(SV&DV)}
which is the $\mathbb{Z}_2$-gauge
extension of a point symmetry group $\mathbf{P}_{\bm{k}}'\subseteq\mathbf{P}_{\bm{k}}$.
\cite{SM:prsym,SM:DPreps(DVtimesDV)}
Similar to electron eigenfunctions, \cite{D2008} spinon eigenfunctions may have essential
degeneracies at special points in the Brillouin zone and these band degeneracies are lifted
as we move away from the high symmetry points [cf. Fig. \ref{F:RamanSquare}(a)].

   The Hilbert space of the spin Hamiltonian (\ref{E:Hspin}) is block-diagonal with respect to
flux configurations $\{W_p,W_c\}$, consisting of $2^{\frac{L}{2}+1}$ blocks of dimension
$2^{\frac{L}{2}-1}\times 2^{\frac{L}{2}-1}$,
while that of the augmented Majorana Hamiltonian (\ref{E:HMajorana}) is block-diagonal
with respect to bond configurations
$
 \{
   u_{<\lambda,\lambda'>}^{\bm{r}_l\to\bm{r}_{l'}}
 \}
$
as well as $\{W_p,W_c\}$, consisting of
$2^{\frac{3}{2}L}$ blocks of dimension $2^{\frac{L}{2}}\times 2^{\frac{L}{2}}$.
Four Majorana fermions at each site have $2^{2L}$ degrees of freedom, containing
``unphysical states" \cite{Y217202,P165414} to be projected out by the operator
$
   \mathcal{P}
  =\prod_{l=1}^{N^2}\prod_{\lambda=1}^q
   \frac{1}{2}
   (1+\eta_{\bm{r}_l:\lambda}^x\eta_{\bm{r}_l:\lambda}^y\eta_{\bm{r}_l:\lambda}^z
    c_{\bm{r}_l:\lambda})
$ \cite{Y217202,P165414,Z014403,U220404}.
We can express $\mathcal{P}$ in terms of the bond operators
$\hat{u}_{<\lambda,\lambda'>}^{\bm{r}_l\to\bm{r}_{l'}}$ and
quasiparticle occupation operators
$
    \alpha_{\bm{k}_\kappa:\lambda}^\dagger\alpha_{\bm{k}_\kappa:\lambda}
$
and therefore straightforwardly apply it to quasiparticle states with given background gauge
fields $\{u_{<\lambda,\lambda'>}^{\bm{r}_l\to\bm{r}_{l'}}\}$.
Physical and unphysical states in each gauge-fixed block of the Hilbert space can be
distinguished according as the number of emergent (complex) fermions in them is even or odd.
Physical states against the ground gauge fields consist of even numbers of quasiparticles
$
    \alpha_{\bm{k}_\kappa:\lambda}^\dagger\alpha_{\bm{k}_\kappa:\lambda}
$.
$W_p$'s of the constituent polygons in the ground state are all
$-1$, $+1$, or either of $+i$ and $-i$ according as their $N_p$'s are
$4l$, $4l+2$, or $2l+1$ with $l\in\mathbb{N}$ (Fig. \ref{F:2DKitaevModels}).
\cite{P134404,M041103}
Apart from the twofold degeneracy due to the constituent triangles,
$W_1=\cdots=W_{\frac{L}{3}}=\pm i$ in Fig. \ref{F:2DKitaevModels}(b),
the ground states of the gauged toruses Figs. \ref{F:2DKitaevModels}(a)--\ref{F:2DKitaevModels}(d)
are all quadruply degenerate due to the topological eigenvalues,
$W_a=\pm 1$ and $W_b=\pm 1$. \cite{H035145}
\begin{figure*}[t]
\centering
\includegraphics[width=177mm]{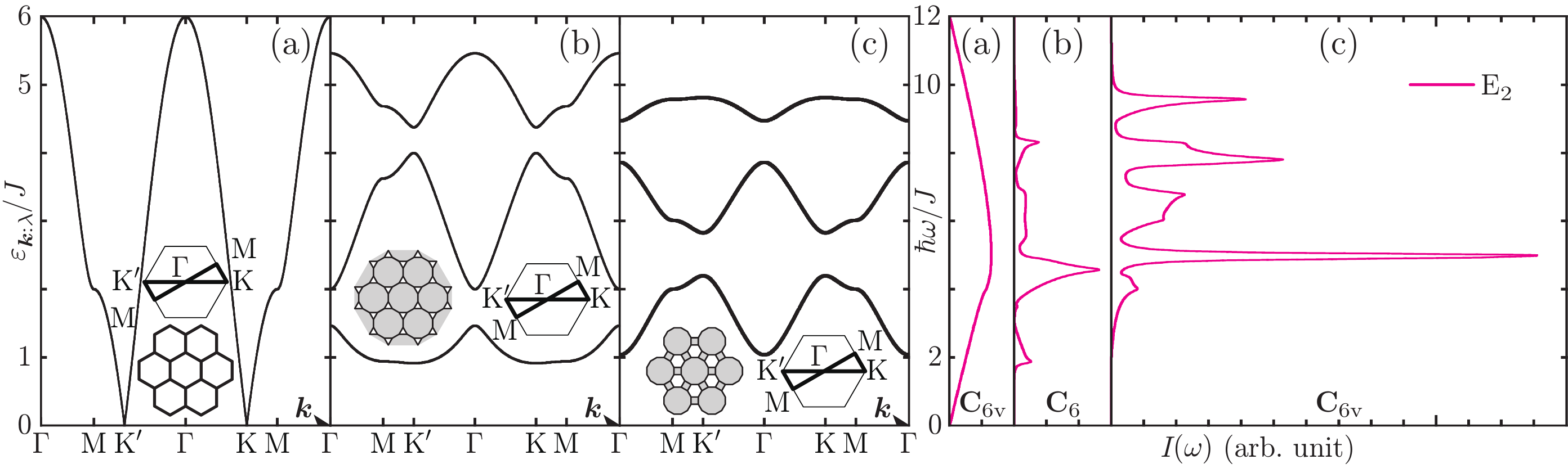}
\vspace*{-5mm}
\caption{(Color online)
         Spinon excitation energies $\varepsilon_{\bm{k}:\lambda}$ $(\lambda=1,\cdots,\frac{q}{2})$
         and Raman intensities $I(\omega)$ of the gauge-ground Kitaev pure (a)-, triangle (b)-, and
         square-hexagon (c)-honeycomb models.
         All the three bands in (c) are doubly degenerate due to
         the primitive-translation-invariant gauged twofold rotational symmetry.
         The Raman responses do not depend on $(\varphi_{\mathrm{in}},\varphi_{\mathrm{sc}})$
         at all.}
\label{F:RamanHoneycomb}
\vspace*{-5mm}
\end{figure*}

   We calculate the intensity of Raman scattering through the LF vertex
\cite{K187201,F514,S1068,S365,P094439} in the ground state $|0\rangle$,
\begin{align}
   &
   I(\omega)
  =\int_{-\infty}^\infty
   \frac{dt\,e^{i\omega t}}{2\pi\hbar L}
   \langle 0|
     e^{\frac{i\mathscr{H}t}{\hbar}}
     \mathcal{R}
     e^{-\frac{i\mathscr{H}t}{\hbar}}
     \mathcal{R}
   |0\rangle;\
   \mathcal{R}
   =\mathcal{R}^{\dagger} 
   \allowdisplaybreaks
   \\
   &
   \equiv
   -J\!\!\!\!\!
   \sum_{<\bm{r}_l:\lambda,\bm{r}_{l'}:\lambda'>}\!\!\!
   (\bm{e}_{\mathrm{in}}
   \cdot
   \bm{d}_{<\lambda,\lambda'>}^{\bm{r}_l\to\bm{r}_{l'}})
   (\bm{e}_{\mathrm{sc}}
   \cdot
   \bm{d}_{<\lambda,\lambda'>}^{\bm{r}_l\to\bm{r}_{l'}})
   \sigma_{\bm{r}_l:\lambda}^{\alpha(\bm{r}_l:\lambda,\bm{r}_{l'}:\lambda')}
   \sigma_{\bm{r}_{l'}:\lambda'}^{\alpha(\bm{r}_l:\lambda,\bm{r}_{l'}:\lambda')}
   \nonumber
   \allowdisplaybreaks
   \\
   &
  =iJ\!\!\!\!\!
   \sum_{<\bm{r}_l:\lambda,\bm{r}_{l'}:\lambda'>}\!\!\!
   (\bm{e}_{\mathrm{in}}
   \cdot
   \bm{d}_{<\lambda,\lambda'>}^{\bm{r}_l\to\bm{r}_{l'}})
   (\bm{e}_{\mathrm{sc}}
   \cdot
   \bm{d}_{<\lambda,\lambda'>}^{\bm{r}_l\to\bm{r}_{l'}})
   \hat{u}_{<\lambda,\lambda'>}
          ^{\bm{r}_l\to\bm{r}_{l'}}
   c_{\bm{r}_l:\lambda}c_{\bm{r}_{l'}:\lambda'},
   \label{E:I(w)LF}
\end{align}
where
$\bm{e}_{\mathrm{in}/\mathrm{sc}}
\equiv(\cos\varphi_{\mathrm{in}/\mathrm{sc}},\sin\varphi_{\mathrm{in}/\mathrm{sc}})$
are the incident and scattered light polarization vectors
and $\bm{d}_{<\lambda,\lambda'>}^{\bm{r}_l\to\bm{r}_{l'}}$ is the lattice vector
$\bm{r}_l:\lambda\to\bm{r}_{l'}:\lambda'$.
When the ground state $|0\rangle$ belongs to the double group $\widetilde{\mathbf{P}}$,
\cite{SM:prsym,SM:Irreps(SV&DV)}
it is useful to decompose the LF vertex $\mathcal{R}$ into irreducible representations
$\widetilde{\varXi}_j$ of $\widetilde{\mathbf{P}}$,
$
   \mathcal{R}
  =\sum_{j}'
   \sum_{\mu=1}^{d_{\widetilde{\varXi}_j}^{\widetilde{\mathbf{P}}}}
   E_{\widetilde{\varXi}_j:\mu}^{\widetilde{\mathbf{P}}}
   \mathcal{R}_{\widetilde{\varXi}_j:\mu}^{\widetilde{\mathbf{P}}}
  =\mathop{{\sum_j}'}
   \sum_{\mu=1}^{d_{\varXi_j}^{\mathbf{P}}}
   E_{\varXi_j:\mu}^{\mathbf{P}}
   \mathcal{R}_{\varXi_j:\mu}^{\mathbf{P}}
$,
where $\sum_{j}'$ runs over LF-active irreducible representations, which are necessarily
\textit{real} and \textit{single-valued},
$d_{\widetilde{\varXi}_j}^{\widetilde{\mathbf{P}}}$ denotes the dimension of
$\widetilde{\varXi}_j$, and
note that all the single-valued irreducible representations of $\widetilde{\mathbf{P}}$ are
nothing but those of the corresponding point symmetry group $\mathbf{P}$. \cite{D175}
When we put
$\mathcal{R}
=\sum_{\beta,\gamma}e_{\mathrm{in}}^\beta e_{\mathrm{sc}}^\gamma\mathcal{R}^{\beta\gamma}$
with
\begin{align}
   \!\!\!
   \mathcal{R}^{\beta\gamma}
  \equiv
   -J\!\!\!\!\!\!\!
   \sum_{<\bm{r}_l:\lambda,\bm{r}_{l'}:\lambda'>}\!\!\!\!\!
   (d_{<\lambda,\lambda'>}^{\bm{r}_l\to\bm{r}_{l'}})^\beta
   (d_{<\lambda,\lambda'>}^{\bm{r}_l\to\bm{r}_{l'}})^\gamma
   \sigma_{\bm{r}_l:\lambda}^{\alpha(\bm{r}_l:\lambda,\bm{r}_{l'}:\lambda')}
   \sigma_{\bm{r}_{l'}:\lambda'}^{\alpha(\bm{r}_l:\lambda,\bm{r}_{l'}:\lambda')},
\end{align}
symmetry-definite LF vertices are given by
\newcommand\StepSubequations{
  \stepcounter{parentequation}
  \gdef\theparentequation{\arabic{parentequation}}
  \setcounter{equation}{0}}
\begin{subequations}
\begin{alignat}{4}
  \sqrt{2}E_{\mathrm{A}_{1}:1}^{\mathbf{C}_{6\mathrm{v}}}
  &
 =\sqrt{2}E_{\mathrm{A}    :1}^{\mathbf{C}_6}
  &
 =\sqrt{2}E_{\mathrm{A}_{1}:1}^{\mathbf{C}_{4\mathrm{v}}}
  &
 =e_{\mathrm{in}}^{x}e_{\mathrm{sc}}^{x}
  +e_{\mathrm{in}}^{y}e_{\mathrm{sc}}^{y},
  \label{E:E(A1)}
  \\
  \sqrt{2}E_{\mathrm{E}_{2}:1}^{\mathbf{C}_{6\mathrm{v}}}
  &
 =\sqrt{2}E_{\mathrm{E}_{2}:1}^{\mathbf{C}_6}
  &
 =\sqrt{2}E_{\mathrm{B}_{1}:1}^{\mathbf{C}_{4\mathrm{v}}}
  &
 =e_{\mathrm{in}}^{x}e_{\mathrm{sc}}^{x}
  -e_{\mathrm{in}}^{y}e_{\mathrm{sc}}^{y},
  \label{E:E(E2:1)}
  \\
  \sqrt{2}E_{\mathrm{E_{2}}:2}^{\mathbf{C}_{6\mathrm{v}}}
  &
 =\sqrt{2}E_{\mathrm{E}_{2}:2}^{\mathbf{C}_6}
  &
 =\sqrt{2}E_{\mathrm{B}_{2}:1}^{\mathbf{C}_{4\mathrm{v}}}
  &
 =e_{\mathrm{in}}^{x}e_{\mathrm{sc}}^{y}
  +e_{\mathrm{in}}^{y}e_{\mathrm{sc}}^{x},
  \label{E:E(E2:2)}
  \\
  \StepSubequations
  \sqrt{2}\mathcal{R}_{\mathrm{A_{1}}:1}^{\mathbf{C}_{6\mathrm{v}}}
  &
 =\sqrt{2}\mathcal{R}_{\mathrm{A    }:1}^{\mathbf{C}_6}
  &
 =\sqrt{2}\mathcal{R}_{\mathrm{A}_{1}:1}^{\mathbf{C}_{4\mathrm{v}}}
  &
 =\mathcal{R}^{xx}+\mathcal{R}^{yy},
  \label{E:R(A1)}
  \\
  \sqrt{2}\mathcal{R}_{\mathrm{E}_{2}:1}^{\mathbf{C}_{6\mathrm{v}}}
  &
 =\sqrt{2}\mathcal{R}_{\mathrm{E}_{2}:1}^{\mathbf{C}_6}
  &
 =\sqrt{2}\mathcal{R}_{\mathrm{B}_{1}:1}^{\mathbf{C}_{4\mathrm{v}}}
  &
 =\mathcal{R}^{xx}-\mathcal{R}^{yy},
  \label{E:R(E2:1)}
  \\
  \sqrt{2}\mathcal{R}_{\mathrm{E}_{2}:2}^{\mathbf{C}_{6\mathrm{v}}}
  &
 =\sqrt{2}\mathcal{R}_{\mathrm{E}_{2}:2}^{\mathbf{C}_6}
  &
 =\sqrt{2}\mathcal{R}_{\mathrm{B}_{2}:1}^{\mathbf{C}_{4\mathrm{v}}}
  &
 =\mathcal{R}^{xy}+\mathcal{R}^{yx}.
  \label{E:R(E2:2)}
\end{alignat}
\end{subequations}
The identity representations (\ref{E:E(A1)}) and (\ref{E:R(A1)}) commute
with their Hamiltonians (\ref{E:Hspin}), resulting in Rayleigh scattering.
Having in mind that
$
   \langle 0|
   \mathcal{R}_{\varXi_j:\mu}^{\mathbf{P}}
   \mathcal{R}_{\varXi_{j'}:\mu'}^{\mathbf{P}}
   |0\rangle
  =\delta_{jj'}\delta_{\mu\mu'}
   \langle 0|
   \mathcal{R}_{\varXi_j:\mu}^{\mathbf{P}}
   \mathcal{R}_{\varXi_j:\mu}^{\mathbf{P}}
   |0\rangle
$ \cite{D175}
and
$
\langle 0|
\mathcal{R}_{\varXi_j:\mu}^{\mathbf{P}}
\mathcal{R}_{\varXi_j:\mu}^{\mathbf{P}}
|0\rangle
$ no longer depends on $\mu$ with $|0\rangle$ being invariant under every symmetry operation of
$\mathbf{P}$, \cite{P094439}
we find that the symmetry species $\varXi_j$ of $I(\omega)$ should be mediated by
spinon geminate excitations containing the same representation $\varXi_j$,
\begin{align}
   &
   I(\omega)
  ={\sum_j}'
   \sum_{\mu=1}^{d_{\varXi_j}^{\mathbf{P}}}
   I_{\varXi_j:\mu}^{\mathbf{P}}(\omega)
   \left(E_{\varXi_j:\mu}^{\mathbf{P}}\right)^2
  ={\sum_j}'
   I_{\varXi_j:1}^{\mathbf{P}}(\omega)
   \sum_{\mu=1}^{d_{\varXi_j}^{\mathbf{P}}}
   \left(E_{\varXi_j:\mu}^{\mathbf{P}}\right)^2;
   \nonumber
   \allowdisplaybreaks
   \\[-2mm]
   &
   I_{\varXi_j:\mu}^{\mathbf{P}}(\omega)
  \equiv
   \int_{-\infty}^\infty
   \frac{dt\,e^{i\omega t}}{2\pi\hbar L}
   \langle 0|
   e^{\frac{i\mathscr{H}t}{\hbar}}
   \mathcal{R}_{\varXi_j:\mu}^{\mathbf{P}}
   e^{-\frac{i\mathscr{H}t}{\hbar}}
   \mathcal{R}_{\varXi_j:\mu}^{\mathbf{P}}
   |0\rangle
  =\frac{1}{L}
   \sum_{\kappa=1}^{N^2}\sum_{\lambda,\lambda'=1}^{q/2}
   \nonumber
   \allowdisplaybreaks
   \\[-2mm]
   &\qquad\times
   \left|
    \langle 0|
       \alpha_{\bm{k}_\kappa:\lambda}\alpha_{-\bm{k}_\kappa:\lambda'}
       \mathcal{R}_{\varXi_j:\mu}^{\mathbf{P}}
    |0\rangle
   \right|^{2}
   \delta(\hbar\omega-\varepsilon_{\bm{k}_\kappa:\lambda}-\varepsilon_{-\bm{k}_\kappa:\lambda'}).
   \label{E:DecomposedIw}
\end{align}

   How many Raman-active modes are possible in the lattice geometry is most decisive of
whether and how the intensity depends on the light polarization.
Since the pure and decorated honeycomb Kitaev QSLs of triangular geometry,
Figs. \ref{F:2DKitaevModels}(a)--\ref{F:2DKitaevModels}(c),
have one and only Raman-active mode $\mathrm{E}_{2}$ of dimensionality two with
in-plane basis functions such that
\vspace*{-1mm}
\begin{align}
   &
   \left(E_{\mathrm{E}_{2}:1}^{\mathbf{C}_{6\mathrm{v}}}\right)^2
  =\left(E_{\mathrm{E}_{2}:1}^{\mathbf{C}_6            }\right)^2
  =\frac{\cos^2(\varphi_{\mathrm{in}}+\varphi_{\mathrm{sc}})}{2},
   \nonumber
   \allowdisplaybreaks
   \\
   &
   \left(E_{\mathrm{E}_{2}:2}^{\mathbf{C}_{6\mathrm{v}}}\right)^2
  =\left(E_{\mathrm{E}_{2}:2}^{\mathbf{C}_6            }\right)^2
  =\frac{\sin^2(\varphi_{\mathrm{in}}+\varphi_{\mathrm{sc}})}{2},
   \label{E:E(E2)phi}
\end{align}
their Raman responses $I(\omega)$ are completely depolarized (Fig. \ref{F:RamanHoneycomb}),
regardless of further details such as whether or not the ground state
spontaneously breaks time reversal symmetry \cite{Y247203} and
whether the spinon excitation spectrum is gapped or gapless. \cite{Y180404}
Since $(q/2)^2$ varieties of spinon geminate excitations bring spectral weights
at each value of $\bm{k}$, $I(\omega)$ weighs and peaks more and more
from Figs. \ref{F:RamanHoneycomb}(a) to \ref{F:RamanHoneycomb}(c).
In Fig. \ref{F:RamanHoneycomb}(c), the Raman spectrum extends in energy to twice the upper boundary
of the corresponding spinon excitation spectrum ($4.813J$ at $\bm{k}=\mathrm{K},\mathrm{K}'$),
while
in Fig. \ref{F:RamanHoneycomb}(b), that ranges in energy relatively far below twice the upper
boundary of the corresponding spinon excitation spectrum ($5.464J$ at $\bm{k}=\Gamma$).
The highest-lying peak of $I(\omega)$ for the square-hexagon-honeycomb model
Fig. \ref{F:2DKitaevModels}(c) is indeed attributable to the spinon geminate excitations
$\sum_{\lambda,\lambda'=5}^6
 \alpha_{\mathrm{K}:\lambda}^\dagger\alpha_{\mathrm{K}':\lambda'}^\dagger$,
while
that for the triangle-honeycomb model Fig. \ref{F:2DKitaevModels}(b) is mediated by
the spinon geminate excitations
$\sum_{\lambda,\lambda'=2}^3
 \alpha_{\mathrm{K}:\lambda}^\dagger\alpha_{\mathrm{K}':\lambda'}^\dagger$ and
$\alpha_{\mathrm{M}:2}^\dagger\alpha_{\mathrm{M}:3}^\dagger$.
In Fig. \ref{F:RamanHoneycomb}(b), the three spinon excitation bands each are nondegenerate
and the highest-lying one has divergent density of states at
$\bm{k}=\Gamma,\mathrm{M},\mathrm{K},\mathrm{K}'$ with energies
$\varepsilon_{\mathrm{K}:3}=\varepsilon_{\mathrm{K}':3}
<\varepsilon_{\mathrm{M}:3}<\varepsilon_{\Gamma:3}$.
The momentum-canceling points $\mathrm{K}$ and $\mathrm{K}'$ have no reciprocal lattice vector
connecting them and therefore the possible spinon geminate excitation
$\alpha_{\mathrm{K}:3}^\dagger\alpha_{\mathrm{K}':3}^\dagger$ can mediate Raman scattering,
whereas neither $\alpha_{\Gamma:3}^\dagger$ nor $\alpha_{\mathrm{M}:3}^\dagger$
can doubly occur in an attempt to mediate Raman scattering.

   The diamond-square Kitaev QSL of $\widetilde{\mathbf{C}_{4\mathrm{v}}}$ symmetry,
Fig. \ref{F:2DKitaevModels}(d),
has two one-dimensional Raman-active modes $\mathrm{B}_{1}$ and $\mathrm{B}_{2}$.
The basis functions remain the same as (\ref{E:E(E2)phi}), but
$I_{\mathrm{B}_{1}:1}^{\mathbf{C}_{4\mathrm{v}}}(\omega)
\neq
 I_{\mathrm{B}_{2}:1}^{\mathbf{C}_{4\mathrm{v}}}(\omega)$,
yielding the Raman response
$I(\omega)
=I_{\mathrm{B}_{1}:1}^{\mathbf{C}_{4\mathrm{v}}}(\omega)
 \bigl(E_{\mathrm{B}_{1}:1}^{\mathbf{C}_{4\mathrm{v}}}\bigr)^2
+I_{\mathrm{B}_{2}:1}^{\mathbf{C}_{4\mathrm{v}}}(\omega)
 \bigl(E_{\mathrm{B}_{2}:1}^{\mathbf{C}_{4\mathrm{v}}}\bigr)^2$
of strong polarization [Fig. \ref{F:RamanSquare}(b)].
Depolarized light scattering in a liquid state \cite{C172406} may sound plausible and
it is indeed occurrent in frustrated Heisenberg antiferromagnets on
the Kagome \cite{C172406,K024414} and triangular \cite{P174412,V145243} lattices
as well as the honeycomb Kitaev QSL. \cite{K187201}
However, it is the case with an ordered Heisenberg antiferromagnet on
the $\mathbf{C}_{5\mathrm{v}}$ Penrose lattice as well, \cite{Ipreprint}
while the present $\widetilde{\mathbf{C}_{4\mathrm{v}}}$-diamond-square,
$\widetilde{\mathbf{D}_{2\mathrm{h}}}$-harmonic-honeycomb, \cite{P094439,O085101}
and 
$\widetilde{\mathbf{T}}$- and $\widetilde{\mathbf{O}_{\mathrm{h}}}$-polyhedral \cite{Y012003}
Kitaev QSLs exhibit strong polarization in their Raman responses.
One and only multidimensional Raman-active mode, depending on the background lattice geometry
rather than whether liquid or solid, is the key ingredient in depolarization of Raman response.

   Now we consider which symmetry species of
the $\widetilde{\mathbf{C}_{4\mathrm{v}}}$-gauged-lattice \cite{SM:prsym} Raman spectrum
originates in which combination---in the sense of energy, momentum, and symmetry---of
spinon eigenmodes.
We show in Fig. \ref{F:RamanSquare}(a) the spinon dispersion relations of the gauge-ground Kitaev
diamond-square model.
Each spinon eigenmode $\varepsilon_{\bm{k}_\kappa:\lambda}$ belongs to a double-valued irreducible
representation of the double group $\widetilde{\mathbf{P}_{\bm{k}}'}$. \cite{SM:DPreps(DVtimesDV)}
The gauge-ground Majorana Hamiltonian (\ref{E:HMajoranaGSr}) in the real space for
Fig. \ref{F:2DKitaevModels}(d) is invariant to $\widetilde{\mathbf{C}_{4\mathrm{v}}}$,
\cite{SM:prsym}
while each block $\mathcal{H}_{\bm{k}_\kappa}$ of its Fourier transform
(\ref{E:HMajoranaGSk}) in the reciprocal space belongs to any of subsets
$\widetilde{\mathbf{P}_{\bm{k}}'}\subseteq\widetilde{\mathbf{C}_{4\mathrm{v}}}$.
We demand that $\widetilde{\mathbf{P}_{\bm{k}}'}$ should keep $\mathcal{H}_{\bm{k}_\kappa}$
invariant.
Under the present Fourier transformation (\ref{E:FT}),
$\widetilde{\mathbf{P}_{\bm{k}}'}$ should consist of \textit{primitive-translation-invariant}
gauged point symmetry operations, i.e., $\widetilde{\mathbf{P}_{\bm{k}}'}$ may be written as
$\widetilde{\mathbf{P}_{\bm{0}}'}$ with
$\mathbf{P}_{\bm{0}}'\subseteq\mathbf{P}_{\bm{0}}=\mathbf{C}_{4\mathrm{v}}$.
$\widetilde{\mathbf{P}_{\bm{0}}'}$ amounts to $\widetilde{\mathbf{C}_{2\mathrm{v}}}$ even at
the highest symmetry points $\Gamma$ and $\mathrm{M}$ to have one and only double-valued
irreducible representation $\mathrm{E}_{\frac{1}{2}}$ and thus bring doubly degenerate spinon
excitations, as is shown in Fig. \ref{F:RamanSquare}(a).
As we move away from them, $\widetilde{\mathbf{P}_{\bm{0}}'}$ further reduces from
$\widetilde{\mathbf{C}_{2\mathrm{v}}}$ to $\widetilde{\mathbf{C}_{2}}$ or lower and
the two-dimensional real irreducible representation
$\mathrm{E}_{\frac{1}{2}}$ splits into two one-dimensional complex ones
$\mathrm{E}_{\frac{1}{2}}^{(1)}$ and $\mathrm{E}_{\frac{1}{2}}^{(2)}$ to lift the band degeneracy.
Irreducible representations of double groups for the gauged diamond-square (reciprocal) lattice
are detailed in Ref. \citen{SM:Irreps(SV&DV)}.
\begin{table}[t]
\centering
\caption{Direct-product representations made of double-valued irreducible representations
         $\widetilde{\varXi}_i\otimes\widetilde{\varXi}_{i'}$
         and their decompositions into single-valued irreducible representations
         $\widetilde{\varXi}_j$, which are underlined when they are relevant to Raman scattering,
         for gauged $\bm{k}$-point symmetry groups $\widetilde{\mathbf{P}_{\bm{k}}'}$.}
\begin{tabular}{crcclrccc}
\hline \hline
\rule{0mm}{3mm}
$\widetilde{\mathbf{P}_{\bm{k}}'}$
                                       & \multicolumn{3}{c}
                                         {$\widetilde{\varXi}_i\otimes\widetilde{\varXi}_{i'}$}
                                       & $\bigoplus_j\widetilde{\varXi}_j
                                         =\bigoplus_j\varXi_j$
\\[0.5mm]\hline
\raisebox{-4mm}[0mm][0mm]{
$\widetilde{\mathbf{C}_{2(a)}},
 \widetilde{\mathbf{C}_{2(b)}}$}
\rule{0mm}{3.2mm}
                                       & \multicolumn{3}{c}
                                         {\,\,$\mathrm{E}_{\frac{1}{2}}^{(1)}
                                           \otimes
                                           \mathrm{E}_{\frac{1}{2}}^{(1)},\,\,
                                           \mathrm{E}_{\frac{1}{2}}^{(2)}
                                           \otimes
                                           \mathrm{E}_{\frac{1}{2}}^{(2)}
                                          $
                                         }
                                       & \,$\bigl[\underline{\mathrm{B}}\bigr]$
                                       \\[2mm]
                                       & \multicolumn{3}{c}
                                         {\,\,$\mathrm{E}_{\frac{1}{2}}^{(1)}
                                           \otimes
                                           \mathrm{E}_{\frac{1}{2}}^{(2)},\,\,
                                           \mathrm{E}_{\frac{1}{2}}^{(2)}
                                           \otimes
                                           \mathrm{E}_{\frac{1}{2}}^{(1)}
                                          $
                                         }
                                       & \,\ $\underline{\mathrm{A}}$
\\[2mm]\hline
$\widetilde{\mathbf{C}_{2(z)}}$        \rule{0mm}{3.2mm}
                                       & \multicolumn{3}{c}
                                         {\,\,$\mathrm{E}_{\frac{1}{2}}^{(1)}
                                           \otimes
                                           \mathrm{E}_{\frac{1}{2}}^{(2)},\,\,
                                           \mathrm{E}_{\frac{1}{2}}^{(2)}
                                           \otimes
                                           \mathrm{E}_{\frac{1}{2}}^{(1)}
                                          $
                                         }
                                       & \,\ $\underline{\mathrm{A}}$
\\[2mm]\hline
$\widetilde{\mathbf{C}_{2(z)}^{\mathrm{v}(ab)}}$\rule{0mm}{3.8mm}
                                       & \multicolumn{3}{c}
                                         {$\mathrm{E}_{\frac{1}{2}}
                                           \otimes
                                           \mathrm{E}_{\frac{1}{2}}
                                          $
                                         }
                                       & $\bigl\{\underline{\mathrm{A}_{1}}\bigr\}\oplus
                                          \bigl[\underline{\mathrm{A}_{2}}\bigr]\oplus
                                          \bigl[\mathrm{B}_{1}\bigr]\oplus
                                          \bigl[\mathrm{B}_{2}\bigr]$
\\[1.3mm]\hline\hline
\end{tabular}
\label{T:DPRintoIrrep}
\vspace*{-5mm}
\end{table}
\begin{table}[t]
\centering
\caption{Compatibility relations between irreducible representations of
         $\mathbf{C}_{4\mathrm{v}}$ and those of its subgroups $\mathbf{C}_{2\mathrm{v}}$ and
         $\mathbf{C}_2$.
         The gauged diamond-square lattice Fig. \ref{F:2DKitaevModels}(d) is invariant to
         $\widetilde{\mathbf{C}_{4\mathrm{v}}}$, \cite{SM:prsym}
         while each block $\mathcal{H}_{\bm{k}_\kappa}$ of the gauge-ground Majorana Hamiltonian
         (\ref{E:HMajoranaGSr}) belongs to any of subgroups
         $\widetilde{\mathbf{P}_{\bm{k}}'}\subseteq\widetilde{\mathbf{C}_{4\mathrm{v}}}$.
         \cite{SM:DPreps(DVtimesDV)}
         The Raman-active modes of $\mathbf{C}_{4\mathrm{v}}$ are underlined and their
         compatible symmetry species of $\mathbf{P}_{\bm{k}}'$ determine
         \textit{which type of spinon geminate excitations is relevant to which symmetry
         species of the Raman scattering intensities}.}
\begin{tabular}{cccccccc}\hline\hline
  \rule{0mm}{3mm}
  $\mathbf{C}_{4\mathrm{v}}$
  & $\mathrm{A}_{1}$                     & $\mathrm{A}_{2}$
  & $\underline{\mathrm{B}_{1}}$         & $\underline{\mathrm{B}_{2}}$
  & $\mathrm{E}$
  \\[0.8mm] \hline
  \rule{0mm}{3mm}
  $\mathbf{C}_{2(z)}^{\mathrm{v}(ab)}$
  & $\mathrm{A}_{1}$                     & $\mathrm{A}_{2}$
  & $\mathrm{A}_{2}$                     & $\mathrm{A}_{1}$
  & $\mathrm{B}_{1}\oplus\mathrm{B}_{2}$ 
  \\
  $\mathbf{C}_{2(z)}$
  & $\mathrm{A}$                         & $\mathrm{A}$
  & $\mathrm{A}$                         & $\mathrm{A}$
  & $2\mathrm{B}$
  \\
  $\mathbf{C}_{2(a)},\mathbf{C}_{2(b)}$
  & $\mathrm{A}$                         & $\mathrm{B}$
  & $\mathrm{B}$                         & $\mathrm{A}$
  & $\mathrm{A}\oplus\mathrm{B}$
  \\ \hline \hline
\end{tabular}
\label{T:Compatibility}
\vspace*{-5mm}
\end{table}

   Every LF scattering is mediated by a momentum-locked spinon geminate excitation
$\alpha_{\bm{k}_\kappa:\lambda}^\dagger\alpha_{-\bm{k}_\kappa:\lambda'}^\dagger$
and characterized by its direct-product representation made of double-valued irreducible
representations $\widetilde{\varXi}_i$ and $\widetilde{\varXi}_{i'}$
of the double group $\widetilde{\mathbf{P}_{\bm{0}}'}$. \cite{SM:prsym}
Direct-product representations of a nonabelian group are not necessarily irreducible, even though
the constituent representations are irreducible.
Those relevant to spinons in pair with wavevectors $\pm\bm{k}_\kappa$ decompose
into single-valued irreducible representations of $\widetilde{\mathbf{P}_{\bm{0}}'}$,
as is shown in Table \ref{T:DPRintoIrrep} and illustrated in more detail in
Ref. \citen{SM:DPreps(DVtimesDV)}.
Every single-valued irreducible representation of $\widetilde{\mathbf{P}_{\bm{0}}'}$ remains
the same as that of the corresponding point symmetry group $\mathbf{P}_{\bm{0}}'$,
$\widetilde{\varXi}_i\otimes\widetilde{\varXi}_{i'}
=\bigoplus_j\widetilde{{\varXi}_j}
=\bigoplus_j            \varXi_j$.
Every direct product of the two same representations reads a sum of symmetric and/or antisymmetric
representations,
$\widetilde{\varXi}_i\otimes\widetilde{\varXi}_i
\equiv
 \bigl[\widetilde{\varXi}_i\otimes\widetilde{\varXi}_i\bigr]
 \oplus
 \bigl\{\widetilde{\varXi}_i\otimes\widetilde{\varXi}_i\bigr\}
=\bigoplus_j\bigl[\varXi_j\bigr]
 \bigoplus_{j'}\bigl\{\varXi_{j'}\bigr\}$.
The fermionic geminate excitations
$\alpha_{\Gamma:1}^\dagger\alpha_{\Gamma:2}^\dagger$ and
$\alpha_{\mathrm{M}:1}^\dagger\alpha_{\mathrm{M}:2}^\dagger$
should have antisymmetric representations.
Thus and thus, along high symmetry points in the Brillouin zone,
we can reveal the symmetry species of spinon geminate excitations
$\alpha_{\bm{k}:\lambda}^\dagger\alpha_{-\bm{k}:\lambda'}^\dagger$,
each compatible with one or more of irreducible representations of the full point symmetry group
$\mathbf{C}_{4\mathrm{v}}$ (Table \ref{T:Compatibility}) and
bringing spectral weights $I(\bm{k},-\bm{k};\omega)$ [Fig. \ref{F:RamanSquare}(c)]
when their compatible symmetry mode(s) of $\mathbf{C}_{4\mathrm{v}}$ are Raman active.
The Raman-active modes of the $\widetilde{\mathbf{C}_{4\mathrm{v}}}$ gauged lattice are
$\mathrm{B_1}$ and $\mathrm{B_2}$, selectively available from
$(\varphi_{\mathrm{in}},\varphi_{\mathrm{sc}})=(0,0)$ and $(0,\frac{\pi}{2})$, respectively.
Suppose we consider two spinons
$\alpha_{(k,k)/\sqrt{2}:\lambda}^\dagger$ and
$\alpha_{-(k,k)/\sqrt{2}:\lambda'}^\dagger$
$(\lambda,\lambda'=1,2)$ on the ways from $\Gamma$ to $\mathrm{M}$
of $\widetilde{\mathbf{C}_{2(a)}}$ symmetry.
The $\mathrm{E}_{\frac{1}{2}}^{(\lambda)}\otimes\mathrm{E}_{\frac{1}{2}}^{(\lambda)}$ pairs
belong to the symmetry species $\mathrm{B}$ of $\mathbf{C}_{2(a)}$, compatible with
LF-Raman-inactive $\mathrm{A}_2$ and Raman-active $\mathrm{B}_1$ of
$\mathbf{C}_{4\mathrm{v}}$,
and therefore bring spectral weights
$\sum_{\lambda=1}^2
 I\bigl[
   \bm{k},-\bm{k};
   (\varepsilon_{\bm{k}:\lambda}+\varepsilon_{-\bm{k}:\lambda})/\hbar
  \bigr]
\equiv
2I\bigl[
   \bm{k};
   (\varepsilon_{\bm{k}:1}+\varepsilon_{\bm{k}:2})/\hbar
  \bigr]$
at $(\varphi_{\mathrm{in}},\varphi_{\mathrm{sc}})=(0,0)$.
The $\mathrm{E}_{\frac{1}{2}}^{(\lambda)}\otimes\mathrm{E}_{\frac{1}{2}}^{(3-\lambda)}$ pairs
belong to the symmetry species $\mathrm{A}$ of $\mathbf{C}_{2(a)}$, compatible with
LF-Raman-inactive (LF-Rayleigh) $\mathrm{A}_1$ and Raman-active $\mathrm{B}_2$ of
$\mathbf{C}_{4\mathrm{v}}$,
and therefore bring spectral weights
$\sum_{\lambda=1}^2
 I\bigl[
   \bm{k},-\bm{k};
   (\varepsilon_{\bm{k}:\lambda}+\varepsilon_{-\bm{k}:3-\lambda})/\hbar
  \bigr]
\equiv
 I\bigl(
   \bm{k};
   2\varepsilon_{\bm{k}:1}/\hbar
  \bigr)
+I\bigl(
   \bm{k};
  2\varepsilon_{\bm{k}:2}/\hbar
  \bigr)$
at $(\varphi_{\mathrm{in}},\varphi_{\mathrm{sc}})=(0,\frac{\pi}{2})$, where single spinon
excitation modes $\varepsilon_{\bm{k}:1}$ and $\varepsilon_{\bm{k}:2}$ each separately appear.

   The key ingredient of depolarized Raman response is one and only multidimensional Raman-active
mode of geometric origin.
It occurs in honeycomb Heisenberg antiferromagnets without any frustration,
while it breaks down in Kitaev QSLs in nontriangular geometry. \cite{P094439,Y012003}
\textit{Polarized} Raman spectra of Kitaev QSLs serve to identify their single spinon excitation
modes as functions of their momenta, even though the total momentum of each pair of mediating
spinons is locked to zero.
Polarized photons can distinguish between spinon geminate excitations of different symmetries
and therefore reveal the projective symmetries of their constituent spinons.
While the depolarization of Raman response in Kitaev QSLs trivially breaks down with
higher-order vertices under increasing itinerancy and decreasing correlation,
the breakdown is possible within the solvable Hamiltonian and LF scheme.
On the one hand nontriangular geometries are realizable on honeycomb lattices
with site dilution \cite{W115146} and/or bond disorder, \cite{Z014403}
but on the other hand
Kitaev QSLs in the cylinder \cite{G157203} and ribbon \cite{S012046} geometries
are intriguing both theoretically and experimentally and may be ideal
targets of the present approach.

   We thank J. Ohara for useful comments.
This work was supported by
the Ministry of Education, Culture, Sports, Science, and Technology of Japan.

\end{document}


\maketitle
\section*{S1. Projective Symmetry Operations on Gauge-Ground Kitaev Spin Planes}

\begin{figure*}[h!]
\centering
\includegraphics[width=177mm]{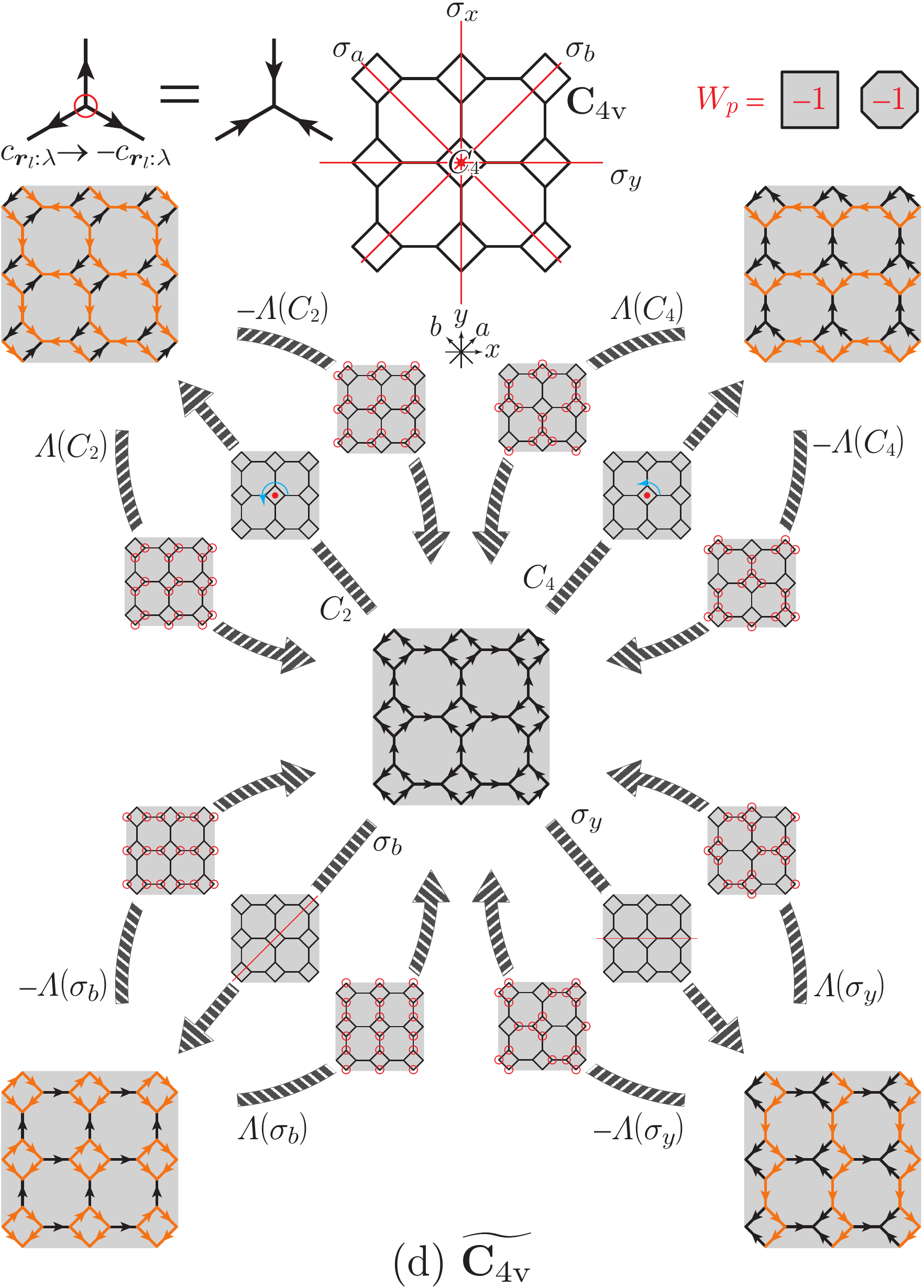}
\caption{Gauged rotations and mirror operations of gauge-ground Kitaev spin planes consisting of
         the pure (a)-, triangle (b)-, square-hexagon (c)-honeycomb and diamond-square (d)
         lattices.
         Mirror operations $\sigma\in\mathbf{C}_{6\mathrm{v}}$ of (b) can be followed by no such
         gauge transformation as to recover the initial bond configuration.
         Rotations and mirror operations
         $C_{4},C_4^{-1},\sigma_x,\sigma_y\in\mathbf{C}_{4\mathrm{v}}$ of (d)
         can be followed by no such gauge transformation as to recover the initial bond
         configuration with the primitive translation vectors remaining unchanged.}
\label{F:prSG}
\end{figure*}

   We discuss the Kitaev Hamiltonian
\begin{align}
   \mathscr{H}
 =-J\sum_{<\bm{r}_l:\lambda,\bm{r}_{l'}:\lambda'>}
   \sigma_{\bm{r}_l:\lambda}^{\alpha(\bm{r}_l:\lambda,\bm{r}_{l'}:\lambda')}
   \sigma_{\bm{r}_{l'}:\lambda'}^{\alpha(\bm{r}_l:\lambda,\bm{r}_{l'}:\lambda')},
   \label{E:Hspin}
\end{align}
on various two-dimensional lattices, where
$(\sigma_{\bm{r}_l:\lambda}^x,\sigma_{\bm{r}_l:\lambda}^y,\sigma_{\bm{r}_l:\lambda}^z)$
$(l=1,\cdots,N^2\equiv L/q;\,\lambda=1,\cdots,q)$ are the Pauli matrices attached to
the $\lambda$th site in the $l$th unit at $\bm{r}_l$ and
$<\bm{r}_l:\lambda,\bm{r}_{l'}:\lambda'>$ runs over $3L/2$ nearest-neighbor bonds.
Denoting the primitive translation vectors in each lattice by $\bm{a}$ and $\bm{b}$,
we adopt a periodic boundary condition, $\bm{r}_l+N\bm{a}=\bm{r}_l+N\bm{b}=\bm{r}_l$.
The number of sites $L$ reads $qN^2$ with
$q=2$ [Fig. \ref{F:prSG}(a)],
$q=6$ [Fig. \ref{F:prSG}(b)],
$q=12$ [Fig. \ref{F:prSG}(c)], and
$q=4$ [Fig. \ref{F:prSG}(d)].

   When we introduce four Majorana fermions at each site through the relation
$\sigma_{\bm{r}_l:\lambda}^\alpha=i\eta_{\bm{r}_l:\lambda}^\alpha c_{\bm{r}_l:\lambda}$
and define nearest-neighbor bond operators as
$
   \hat{u}_{<\lambda,\lambda'>}^{\bm{r}_l\to\bm{r}_{l'}}
  \equiv
  i\eta_{\bm{r}_l:\lambda}^{\alpha(\bm{r}_l:\lambda,\bm{r}_{l'}:\lambda')}
   \eta_{\bm{r}_{l'}:\lambda'}^{\alpha(\bm{r}_l:\lambda,\bm{r}_{l'}:\lambda')}
  =-\hat{u}_{<\lambda',\lambda>}^{\bm{r}_{l'}\to\bm{r}_l}
$,
the Hamiltonian is rewritten into
\begin{align}
   \mathscr{H}
  =iJ\sum_{<\bm{r}_l:\lambda,\bm{r}_{l'}:\lambda'>}
   \hat{u}_{<\lambda,\lambda'>}
          ^{\bm{r}_l\to\bm{r}_{l'}}
   c_{\bm{r}_l:\lambda}c_{\bm{r}_{l'}:\lambda'}.
   \label{E:HMajorana}
\end{align}
Every $\hat{u}_{<\lambda,\lambda'>}^{\bm{r}_l\to\bm{r}_{l'}}$
is a conserved quantity and behaves as a $\mathbb{Z}_2$ classical variable,
$
   u_{<\lambda,\lambda'>}^{\bm{r}_l\to\bm{r}_{l'}}
  =\pm 1
$.
Once we fix the background gauge fields to one of the $2^{\frac{3}{2}L}$ configurations,
the Majorana Hamiltonian (\ref{E:HMajorana}) becomes exactly solvable.
The ground-flux-configuration sector of (\ref{E:HMajorana}) may be written as
\begin{align}
   \mathscr{H}
  =\frac{iJ}{2}
   \sum_{l=1}^{N^2}
   \sum_{\bm{\delta}=\bm{0},\pm\bm{a},\pm\bm{b}}
   \sum_{\lambda=1}^q\sum_{\lambda'=1}^q
   u_{<\lambda,\lambda'>}^{\bm{\delta}}
   c_{\bm{r}_l:\lambda}c_{\bm{r}_l+\bm{\delta}:\lambda'},
   \label{E:HMajoranaGSr}
\end{align}
where
$
   u_{<\lambda,\lambda'>}^{\bm{\delta}}
  \equiv
   u_{<\lambda,\lambda'>}^{\bm{r}_l\to\bm{r}_l+\bm{\delta}}
$
no longer depend on the position $\bm{r}_l$.
We consider gauged point symmetry operations \cite{M041103,W165113} on
the gauge-ground Majorana Hamiltonian (\ref{E:HMajoranaGSr})
[Figs. \ref{F:prSG}(a)--\ref{F:prSG}(d)].
Carrying out the Fourier transformation
\vspace*{-1mm}
\begin{align}
   \gamma_{\bm{k}_\kappa:\lambda}
  =\frac{1}{\sqrt{2}N}\sum_{l=1}^{N^2}
   e^{i\bm{k}_\kappa\cdot\bm{r}_l}
   c_{\bm{r}_l:\lambda},\ 
   c_{\bm{r}_l:\lambda}
  =\frac{\sqrt{2}}{N}\sum_{\kappa=1}^{N^2}
   e^{-i\bm{k}_\kappa\cdot\bm{r}_l}
   \gamma_{\bm{k}_\kappa:\lambda}
   \label{E:FT}
\end{align}
with $\gamma_{-\bm{k}_\kappa:\lambda}=\gamma_{\bm{k}_\kappa:\lambda}^\dagger$ in mind yields
\begin{align}
   &
   \mathscr{H}
  =iJ\sum_{\kappa=1}^{N^2}
   \sum_{\lambda=1}^q\sum_{\lambda'=1}^q
   u_{<\lambda,\lambda'>}(\bm{k}_\kappa)
   \gamma_{\bm{k}_\kappa:\lambda}^\dagger\gamma_{\bm{k}_\kappa:\lambda'};
   \nonumber
   \allowdisplaybreaks
   \\
   &
   u_{<\lambda,\lambda'>}(\bm{k}_\kappa)
  \equiv
   \sum_{\bm{\delta}=\bm{0},\pm\bm{a},\pm\bm{b}}
   u_{<\lambda,\lambda'>}^{\bm{\delta}}
   e^{-i\bm{k}_\kappa\cdot\bm{\delta}}
  =u_{<\lambda,\lambda'>}^*(-\bm{k}_\kappa).
   \label{E:HMajoranaGSk}
\end{align}
The flux operator defined for an $N_p$-sided polygon $p$
\begin{align}
   \hat{W}_p
   &
  \equiv
   \prod_{<\bm{r}_l:\lambda,\bm{r}_{l'}:\lambda'>\in\partial p}
   \sigma_{\bm{r}_l:\lambda}^{\alpha(\bm{r}_l:\lambda,\bm{r}_{l'}:\lambda')}
   \sigma_{\bm{r}_{l'}:\lambda'}^{\alpha(\bm{r}_{l'}:\lambda,\bm{r}_{l'}:\lambda')}
   \nonumber \\
   &
  =(-i)^{N_p}
   \prod_{<\bm{r}_l:\lambda,\bm{r}_{l'}:\lambda'>\in\partial p}
   \hat{u}_{<\lambda,\lambda'>}^{\bm{r}_l\to\bm{r}_{l'}}
   \label{E:Wp}
\end{align}
also behaves as a classical variable,
$W_p=\pm 1$ or $\pm i$ according as $N_p$ is even or odd.
The eigenspectrum of (\ref{E:HMajoranaGSr}) depends on the bond configuration
$\{u_{<\lambda,\lambda'>}^{\bm{r}_l\to\bm{r}_{l'}}\}$ only through the flux configuration
$\{W_p\}$.
The ground-state flux configuration is such that \cite{P134404}
all $W_p$'s are $-1$, $+1$, or either of $+i$ and $-i$ according as their $N_p$'s are
$4l$, $4l+2$, or $2l+1$ with $l\in\mathbb{N}$ [cf. Figs. \ref{F:prSG}(a)--\ref{F:prSG}(d)].

   The pure-, triangle-, and square-hexagon-honeycomb lattices are of $\mathbf{C}_{6\mathrm{v}}$
point symmetry [\ref{F:prSG}(a)--\ref{F:prSG}(c)], while the diamond-square lattice is of
$\mathbf{C}_{4\mathrm{v}}$ point symmetry  [\ref{F:prSG}(d)].
Once these lattices are gauged in the context of the Kitaev quantum spin liquid, they no longer
belong to their original point symmetry groups $\mathbf{P}_{\rm org}$ but may become invariant
under gauged point symmetry operations $\widetilde{P}\in\widetilde{\mathbf{P}}$ with
$\mathbf{P}\subseteq\mathbf{P}_{\rm org}$.
Note that $\mathbf{P}$ is not necessarily equal to $\mathbf{P}_{\rm org}$.

   Let us start with a simple example.
Figure \ref{F:prSG}(a) illustrates gauged point symmetry operations on the gauge-ground Kitaev
honeycomb Hamiltonian.
Suppose we rotate it by $\frac{\pi}{3}$ about the normal vector, denoted by $C_6$,
and then gauge some Majorana fermions as
$c_{\bm{r}_l:\lambda}\to -c_{\bm{r}_l:\lambda}$, or equivalently, change the signs of their
relevant bonds as
$u_{<\lambda,\lambda'>}^{\bm{r}_l\to\bm{r}_{l'}}
\to
-u_{<\lambda,\lambda'>}^{\bm{r}_l\to\bm{r}_{l'}}$,
so as to recover the initial bond configuration.
Any local gauge transformation $c_{\bm{r}_l:\lambda}\to -c_{\bm{r}_l:\lambda}$ reverses its
three surrounding bond variables.
Given a point symmetry operation $P\in\mathbf{C}_{6\mathrm{v}}$, there are two such local gauge
transformations, denoted by $\pm\varLambda(P)$.
Let us abbreviate a couple of these serial transformations as
$+\varLambda(P)P\equiv\overline{P}$
and
$-\varLambda(P)P\equiv\underline{P}$
and denote them unifiedly as $\widetilde{P}$.
We find that the gauged honeycomb belongs to the double group
$\widetilde{\mathbf{C}_{6\mathrm{v}}}$ and therefore $\mathbf{P}=\mathbf{P}_{\mathrm{org}}$
in this case.

   Figure \ref{F:prSG}(b) illustrates mirror operations as well as gauged rotations on
the gauge-ground Kitaev triangle-honeycomb Hamiltonian.
Every mirror operation reverses all the flux variables $W_p$ of the constituent triangles,
while any local gauge transformation $c_{\bm{r}_l:\lambda}\to -c_{\bm{r}_l:\lambda}$ results
in reversing the signs of bonds in pair in the three surrounding polygons to keep their flux
variables $W_p$ unchanged.
We find that the symmetry group of the gauged triangle honeycomb is not
$\widetilde{\mathbf{C}_{6\mathrm{v}}}$ but $\widetilde{\mathbf{C}_6}$.
Note that $\mathbf{P}\neq\mathbf{P}_{\mathrm{org}}$ in this case.
In the same manner, we find that
$\mathbf{P}=\mathbf{P}_{\mathrm{org}}=\mathbf{C}_{6\mathrm{v}}$
for the gauged square-hexagon honeycomb and
$\mathbf{P}=\mathbf{P}_{\mathrm{org}}=\mathbf{C}_{4\mathrm{v}}$
for the gauged diamond square.
\clearpage
\begin{figure*}[h!]
\centering
\includegraphics[width=177mm]{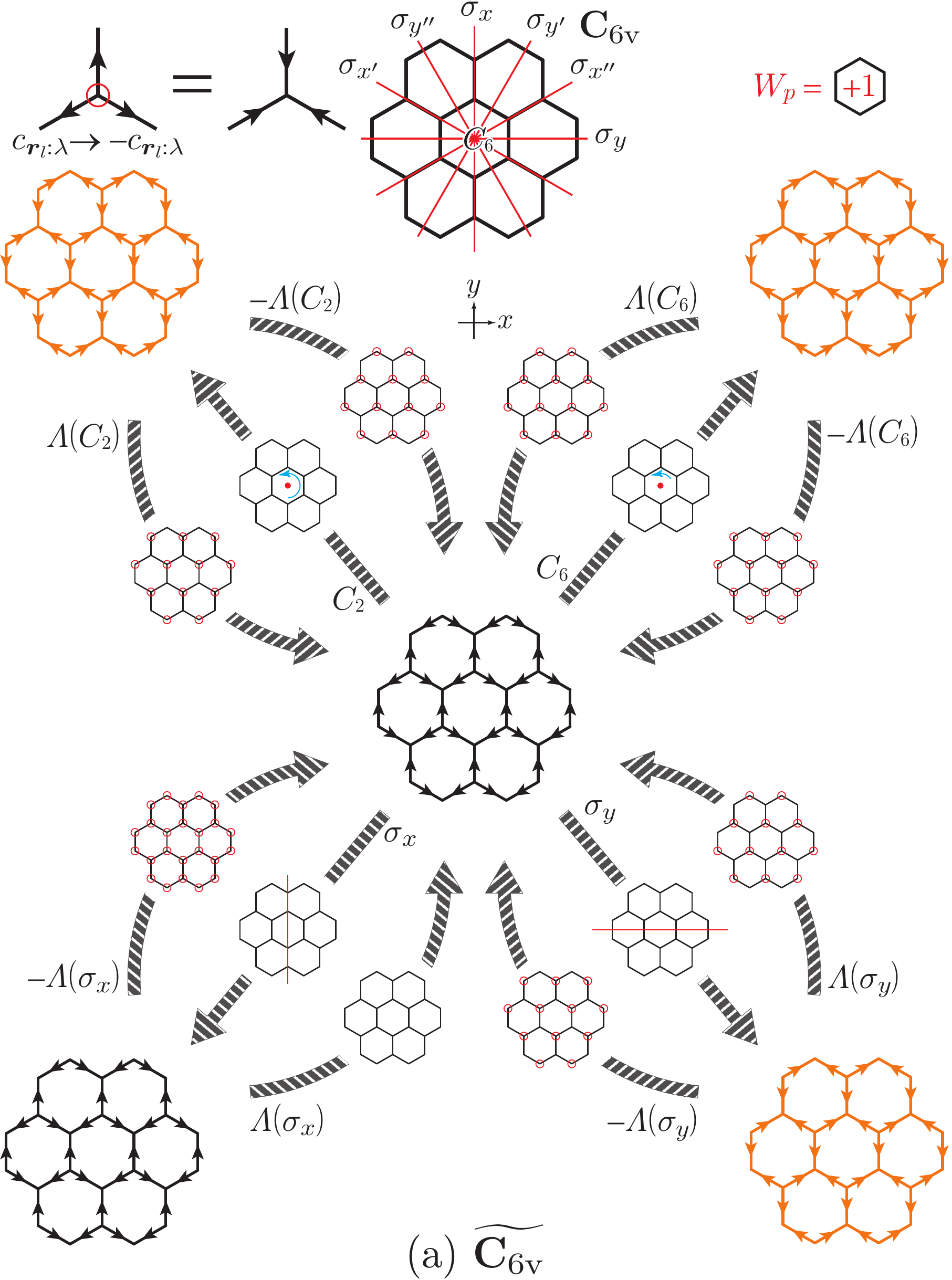}
\vspace{12mm}
\end{figure*}
\begin{figure*}[h!]
\centering
\includegraphics[width=177mm]{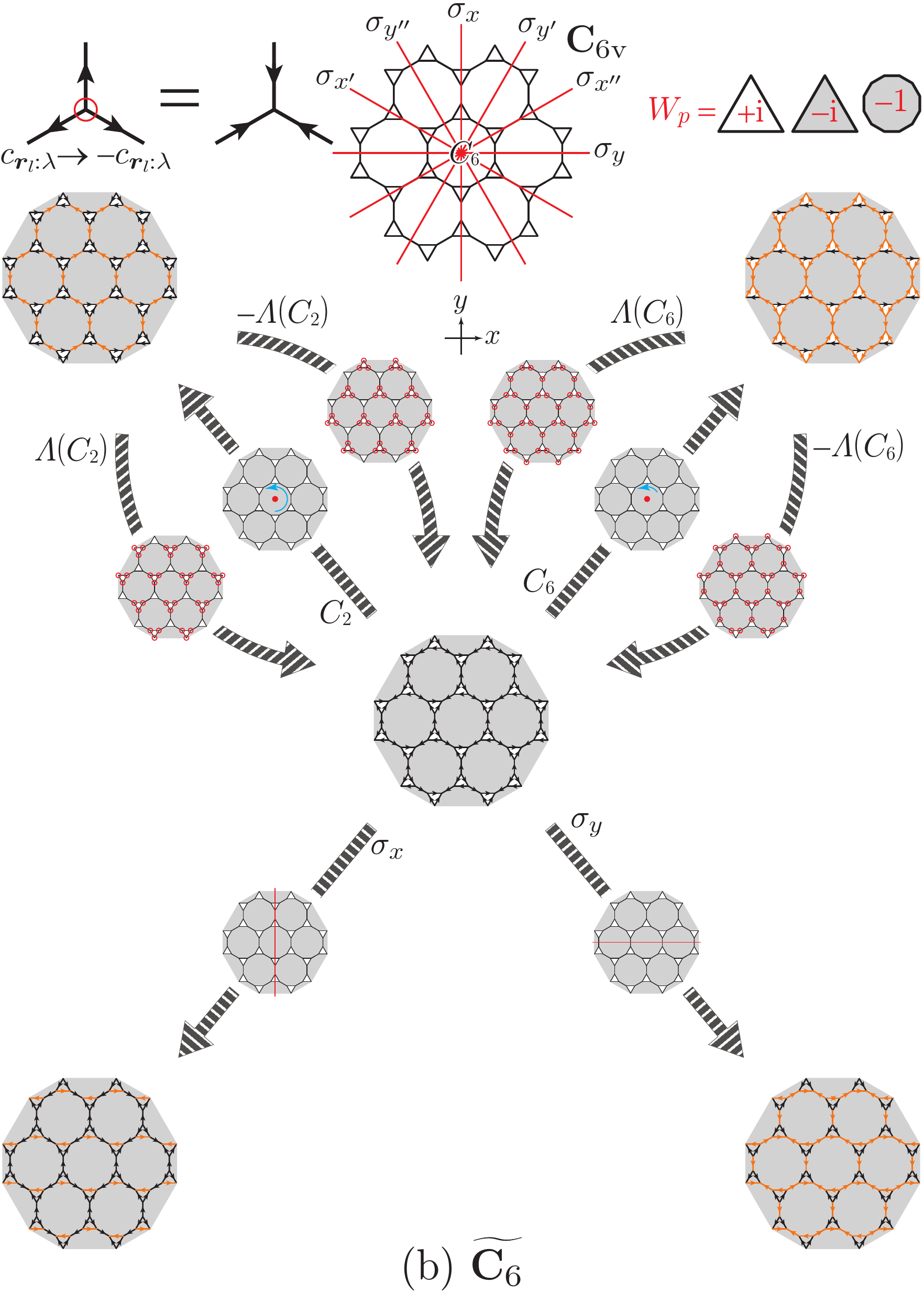}
\vspace{12mm}
\end{figure*}
\begin{figure*}[h!]
\centering
\includegraphics[width=177mm]{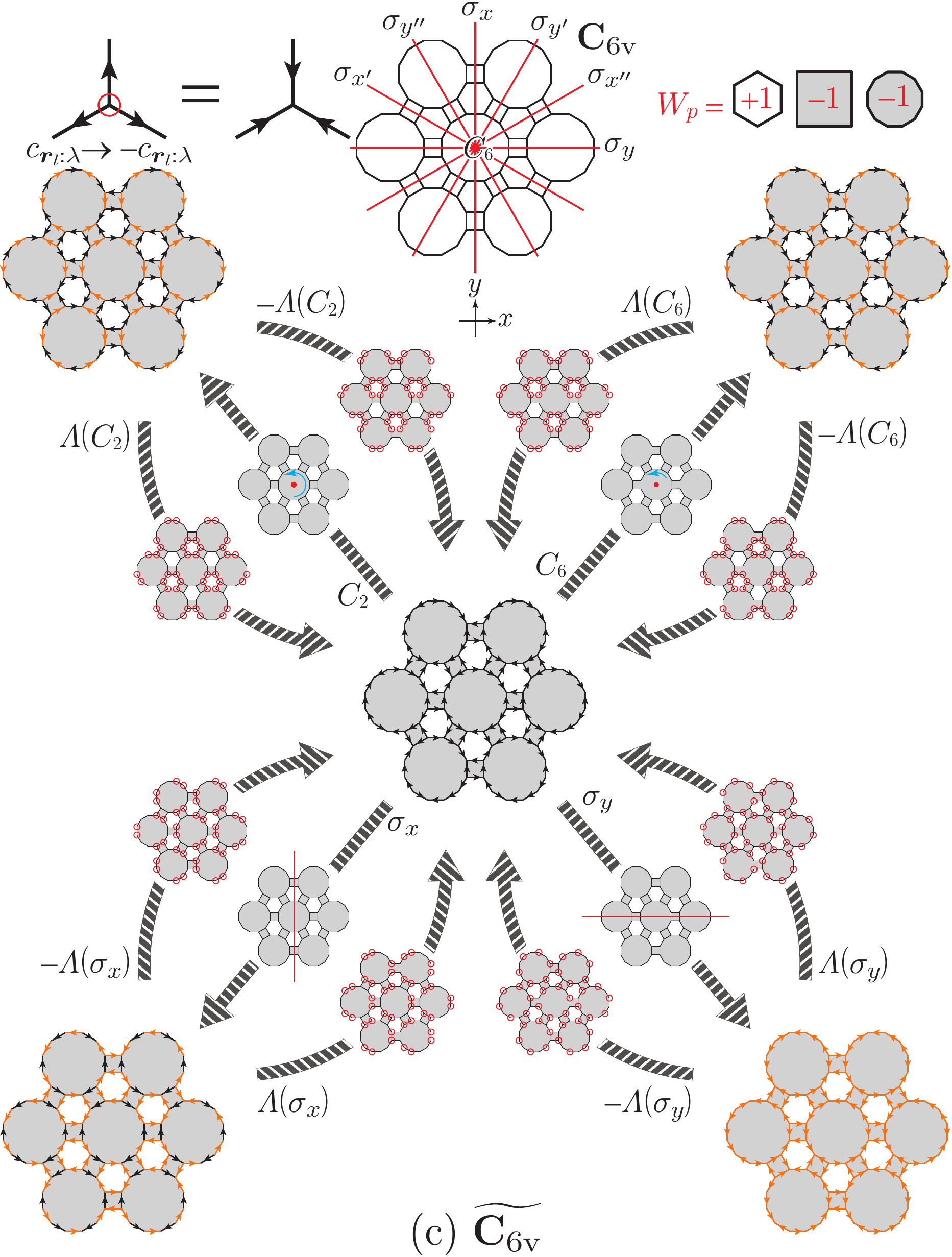}
\vspace{15mm}
\end{figure*}
\clearpage

\section*{S2. Irreducible Representations of Double Groups for
          the Gauged Diamond-Square Lattice}

   We denote the orders of a point symmetry group $\mathbf{P}$ and its double covering group
$\widetilde{\mathbf{P}}$ by $g^{\mathbf{P}}$ and $g^{\widetilde{\mathbf{P}}}$, respectively.
Suppose the double cover $\widetilde{\mathbf{P}}$ to be the $\mathbb{Z}_2$-gauge extension
of $\mathbf{P}\subset\mathrm{O}(3)$.
Two group elements
$\widetilde{P}_{1}\in\widetilde{\mathbf{P}}$ and $\widetilde{P}_{2}\in\widetilde{\mathbf{P}}$
are conjugate when we find such an element $\widetilde{P}\in\widetilde{\mathbf{P}}$ as to
satisfy
\begin{align}
   \widetilde{P}_{2}
  =\widetilde{P}\,\widetilde{P}_{1}\widetilde{P}^{-1}.
\end{align}
Every set of conjugate elements forms a class.
The classes of the double group of $\mathbf{C}_{4\mathrm{v}}$ and its subgroups read
\begin{align*}
   \widetilde{\mathbf{C}_{4\mathrm{v}}}:
   &
   \begin{aligned}
   {}
     &
     \{\overline{E}\},\{\underline{E}\},
     \{2\overline{C_{4}}\},\{2\underline{C_{4}}\},
     \{\overline{C_{2}},\underline{C_{2}}\},
     \\[-2mm]
     &
     \{
       \overline{\sigma_{x}},\overline{\sigma_{y}},
       \underline{\sigma_{x}},\underline{\sigma_{y}}
     \},
     \{\overline{\sigma_{a}},\overline{\sigma_{b}},
       \underline{\sigma_{a}},\underline{\sigma_{b}}
     \};
   \end{aligned}
   \\[-1mm]
   \widetilde{\mathbf{C}_{2\mathrm{v}}}:
   &
   \{\overline{E}\},\{\underline{E}\},
   \{\overline{C_{2}},\underline{C_{2}}\},
   \{\overline{\sigma_{x}},\underline{\sigma_{x}}\},
   \{\overline{\sigma_{y}},\underline{\sigma_{y}}\};
   \\[-1mm]
   \widetilde{\mathbf{C}_{2}}:
   &
   \{\overline{E}\},\{\underline{E}\},
   \{\overline{C_{2}}\},\{\underline{C_{2}}\}.
\end{align*}
Supposing the $q$th class $\mathcal{C}_q$ ($q=1,\cdots,n_{\mathcal{C}}^{\widetilde{\mathbf{P}}}$)
of $\widetilde{\mathbf{P}}$ to consist of $h_q$ elements, it reads
$\{h_q\overline{P}_q\}$,
$\{h_q\underline{P}_q\}$, or
$\{\frac{h_q}{2}\overline{P}_q,\frac{h_q}{2}\underline{P}_q\}$.

   The number of (complex) irreducible representations equals how many classes are in the group.
Since all the single-valued (complex) irreducible representations of $\mathbf{P}$, amounting to
$n_{\mathcal{C}}^{\mathbf{P}}$, remain unchanged in $\widetilde{\mathbf{P}}$, we find
$n_{\mathcal{C}}^{\widetilde{\mathbf{P}}}-n_{\mathcal{C}}^{\mathbf{P}}$
double-valued (complex) irreducible representations in $\widetilde{\mathbf{P}}$.
When we denote the $i$th (complex) irreducible representation of
$\mathbf{P}$ ($\widetilde{\mathbf{P}}$) by $\varXi_i$ ($\widetilde{\varXi}_i$) and
its dimensionality by
$d_{\varXi_i}^{\mathbf{P}}$ ($d_{\widetilde{\varXi}_i}^{\widetilde{\mathbf{P}}}$), we have
\begin{align}
   &
   \sum_{i=1}
       ^{n_{\mathcal{C}}^{\mathbf{C}_{4\mathrm{v}}}\equiv 5}
   \left(d_{\varXi_i}^{\mathbf{C}_{4\mathrm{v}}}\right)^2
  =g^{\mathbf{C}_{4\mathrm{v}}}
  =8,
   \nonumber
   \allowdisplaybreaks
   \\
   & 
   \sum_{i=1}
       ^{n_{\mathcal{C}}^{\widetilde{\mathbf{C}_{4\mathrm{v}}}}\equiv 7}
   \left(d_{\widetilde{\varXi}_i}^{\widetilde{\mathbf{C}_{4\mathrm{v}}}}\right)^2
  =g^{\mathbf{C}_{4\mathrm{v}}}
  +\sum_{i=n_{\mathcal{C}}^{\mathbf{C}_{4\mathrm{v}}}+1}
       ^{n_{\mathcal{C}}^{\widetilde{\mathbf{C}_{4\mathrm{v}}}}\equiv 7}
   \left(d_{\widetilde{\varXi}_i}^{\widetilde{\mathbf{C}_{4\mathrm{v}}}}\right)^2
  =g^{\widetilde{\mathbf{C}_{4\mathrm{v}}}}
  =16;
   \\
   &
   \sum_{i=1}
       ^{n_{\mathcal{C}}^{\mathbf{C}_{2\mathrm{v}}}\equiv 4}
   \left(d_{\varXi_i}^{\mathbf{C}_{2\mathrm{v}}}\right)^2
  =g^{\mathbf{C}_{2\mathrm{v}}}
  =4,
   \nonumber 
   \\
   & 
   \sum_{i=1}
       ^{n_{\mathcal{C}}^{\widetilde{\mathbf{C}_{2\mathrm{v}}}}\equiv 5}
   \left(d_{\widetilde{\varXi}_i}^{\widetilde{\mathbf{C}_{2\mathrm{v}}}}\right)^2
  =g^{\mathbf{C}_{2\mathrm{v}}}
  +\sum_{i=n_{\mathcal{C}}^{\mathbf{C}_{2\mathrm{v}}}+1}
       ^{n_{\mathcal{C}}^{\widetilde{\mathbf{C}_{2\mathrm{v}}}}\equiv 5}
   \left(d_{\widetilde{\varXi}_i}^{\widetilde{\mathbf{C}_{2\mathrm{v}}}}\right)^2
  =g^{\widetilde{\mathbf{C}_{2\mathrm{v}}}}
  =8;
   \\
   &
   \sum_{i=1}
       ^{n_{\mathcal{C}}^{\mathbf{C}_{2}}\equiv 2}
   \left(d_{\varXi_i}^{\mathbf{C}_{2}}\right)^2
  =g^{\mathbf{C}_{2}}
  =2,
   \nonumber
   \\
   & 
   \sum_{i=1}
       ^{n_{\mathcal{C}}^{\widetilde{\mathbf{C}_{2}}}\equiv 4}
   \left(d_{\widetilde{\varXi}_i}^{\widetilde{\mathbf{C}_{2}}}\right)^2
  =g^{\mathbf{C}_{2}}
  +\sum_{i=n_{\mathcal{C}}^{\mathbf{C}_{2}}+1}
       ^{n_{\mathcal{C}}^{\widetilde{\mathbf{C}_{2}}}\equiv 4}
   \left(d_{\widetilde{\varXi}_i}^{\widetilde{\mathbf{C}_{2}}}\right)^2
  =g^{\widetilde{\mathbf{C}_{2}}}
  =4
\end{align}
in an attempt to determine the dimensionalities of the double-valued (complex) irreducible
representations $d_{\widetilde{\varXi}_i}^{\widetilde{\mathbf{P}}}\,
 (i=n_{\mathcal{C}}^{\mathbf{P}}+1,\cdots,n_{\mathcal{C}}^{\widetilde{\mathbf{P}}})$.
The characters of $\widetilde{\varXi}_i$ are such that
\begin{alignat}{2}
   \chi_{\widetilde{\varXi}_{i}}^{\widetilde{\mathbf{P}}}(\overline{P})
   &
  =\ \ \chi_{\widetilde{\varXi}_{i}}^{\widetilde{\mathbf{P}}}(\underline{P})
   &&\ 
   (i=1,\cdots,n_{\mathcal{C}}^{\mathbf{P}}),
   \label{E:ChiSingle}
   \\
   \chi_{\widetilde{\varXi}_{i}}^{\widetilde{\mathbf{P}}}(\overline{P})
   &
  =-\chi_{\widetilde{\varXi}_{i}}^{\widetilde{\mathbf{P}}}(\underline{P})
   &&\ 
   (i=n_{\mathcal{C}}^{\mathbf{P}}+1,\cdots,n_{\mathcal{C}}^{\widetilde{\mathbf{P}}}).
   \label{E:ChiDouble}
\end{alignat}
When $\overline{P}$ and $\underline{P}$ belong to the same class, i.e.,
$
   \chi_{\widetilde{\varXi}_{i}}^{\widetilde{\mathbf{P}}}(\overline{P})
  =\chi_{\widetilde{\varXi}_{i}}^{\widetilde{\mathbf{P}}}(\underline{P})
$,
we immediately find
\begin{align}
   \chi_{\widetilde{\varXi}_{i}}^{\widetilde{\mathbf{P}}}(\overline{P})
   &
  =\chi_{\widetilde{\varXi}_{i}}^{\widetilde{\mathbf{P}}}(\underline{P})
  =0\ 
   (i=n_{\mathcal{C}}^{\mathbf{P}}+1,\cdots,n_{\mathcal{C}}^{\widetilde{\mathbf{P}}}).
   \label{E:ZeroClass}
\end{align}
The character orthogonality theorems of the first and second kinds read \cite{D2008}
\begin{align}
   &
   \sum_{q=1}^{n_{\mathcal{C}}^{\widetilde{\mathbf{P}}}}
   h_q
   \chi_{\widetilde{\varXi}_i}^{\widetilde{\mathbf{P}}}(\mathcal{C}_q)^*
   \chi_{\widetilde{\varXi}_{i'}}^{\widetilde{\mathbf{P}}}(\mathcal{C}_q)
  =g^{\widetilde{\mathbf{P}}}\delta_{ii'},
   \label{E:Orthogonal1}
   \\
   &
   \sum_{i=1}^{n_{\mathcal{C}}^{\widetilde{\mathbf{P}}}}
   \chi_{\widetilde{\varXi}_i}^{\widetilde{\mathbf{P}}}(\mathcal{C}_q)^*
   \chi_{\widetilde{\varXi}_i}^{\widetilde{\mathbf{P}}}(\mathcal{C}_r)
  =\frac{g^{\widetilde{\mathbf{P}}}}{h_q}\delta_{qr}.
   \label{E:Orthogonal2}
\end{align}
When we denote the $h_q$ elements of $\mathcal{C}_q$ distinguishably as
$\{\widetilde{P}_q^{(1)},\cdots,\widetilde{P}_q^{(h_q)}\}$,
we can define structure constants as
\vspace{-3mm}
\begin{align}
   \sum_{i=1}^{h_q}\widetilde{P}_q^{(i)}
   \sum_{i'=1}^{h_r}\widetilde{P}_r^{(i')}
  =\sum_{s=1}^{n_{\mathcal{C}}^{\widetilde{\mathbf{P}}}}
   c_{qr:s}
   \sum_{j=1}^{h_j}\widetilde{P}_s^{(j)}
\end{align}
to have another relation,
\vspace{-3mm}
\begin{align}
   h_q h_r
   \chi_{\widetilde{\varXi}_i}^{\widetilde{\mathbf{P}}}(\mathcal{C}_q)
   \chi_{\widetilde{\varXi}_i}^{\widetilde{\mathbf{P}}}(\mathcal{C}_r)
  =d_{\widetilde{\varXi}_i}^{\widetilde{\mathbf{P}}}
   \sum_{s=1}^{n_{\mathcal{C}}^{\widetilde{\mathbf{P}}}}
   h_s c_{qr:s}
   \chi_{\widetilde{\varXi}_i}^{\widetilde{\mathbf{P}}}(\mathcal{C}_s).
   \label{E:Classmulti}
\end{align}
With Eqs. (\ref{E:ZeroClass}), (\ref{E:Orthogonal1}), (\ref{E:Orthogonal2}), and
(\ref{E:Classmulti}) in mind,
we can obtain characters of both single- and double-valued (complex) irreducible
representations of any double group $\widetilde{\mathbf{P}}$.
We list in Tables \ref{T:characterC4vtilde}--\ref{T:characterC2tilde}
characters of irreducible representations of the double group
$\widetilde{\mathbf{C}_{4\mathrm{v}}}$ and those of its subgroups
$\widetilde{\mathbf{C}_{2\mathrm{v}}}$ and $\widetilde{\mathbf{C}_{2}}$
with particular emphasis on the relation between $\widetilde{\mathbf{P}}$ and $\mathbf{P}$.
\vspace{-5mm}
\begin{table}[htb]
\caption{Irreducible representations of the double group $\widetilde{\mathbf{C}_{4\mathrm{v}}}$
         and their characters.}
\begin{tabular}{cccccccccccccccc}
\cline{3-10}\\[-2.8mm]\cline{3-10}
                         &
                         & \rule{0mm}{6mm}
                         & \hspace{-0.5mm}
                           \raisebox{1.5mm}[0mm][0mm]{$\bigl\{\overline{{E}}\bigr\}$}
                           \hspace{-1.5mm}
                         & \hspace{-1.5mm}
                           \raisebox{1.5mm}[0mm][0mm]{$\bigl\{\underline{{E}}\bigr\}$}
                           \hspace{-0.5mm}
                         & \hspace{-1mm}
                           \raisebox{1.5mm}[0mm][0mm]{$\bigl\{2\overline{{C}_{4}}\bigr\}$}
                           \hspace{-2mm}
                         & \hspace{-2mm}
                           \raisebox{1.5mm}[0mm][0mm]{$\bigl\{2\underline{{C}_{4}}\bigr\}$}
                           \hspace{-1mm}
                         & \hspace{-2mm}
                           \shortstack{
                           $\bigl\{\overline{{C}_{2}},$ \\[-1mm]
                           $\,\,\,\,\underline{{C}_{2}}\,\bigr\}$}
                           \hspace{-2mm}
                         & \hspace{-2mm}
                           \shortstack{
                           $\bigl\{\overline{\sigma_{x}},\overline{\sigma_{y}},$ \\[-1mm]
                           $\,\,\,\,
                            \underline{\sigma_{x}},\underline{\sigma_{y}}\,\bigr\}$}
                           \hspace{-2mm} 
                         & \hspace{-2mm}
                           \shortstack{
                           $\bigl\{\overline{\sigma_{a}},\overline{\sigma_{b}},$ \\[-1mm]
                           $\,\,\,\,
                              \underline{\sigma_{a}},\underline{\sigma_{b}}\,\bigr\}$}
                           \hspace{-2mm}
\\[1mm]
\cline{3-10}
\hspace{-3mm}
\raisebox
{-12mm}
[0mm][0mm]
{\smash{
 $\widetilde{\mathbf{C}_{4\mathrm{v}}} 
 \!
 \left\{
 \rule{0mm}{14.8mm}
 \right.$}}
\hspace{-5mm}            & \hspace{-2mm}
                           \raisebox
                           {-7mm}
                           [0mm][0mm]
                           {\smash{
                            $\mathbf{C}_{4\mathrm{v}}\!
                             \left\{
                             \rule{0mm}{10mm}
                             \right.$}}
                           \hspace{-5mm}
                         & $\mathrm{A}_{1}$
                         & \multicolumn{2}{c}{$ 1 $\hphantom{$-$}}
                         & \multicolumn{2}{c}{\hphantom{$-$}$ 1 $\hphantom{$-$}}
                         & $ 1 $
                         & \hphantom{$-$}$ 1 $\hphantom{$-$}
                         & \hphantom{$-$}$1$\hphantom{$-$}
\\
                         &
                         & $\mathrm{A}_{2}$
                         & \multicolumn{2}{c}{$ 1 $\hphantom{$-$}}
                         & \multicolumn{2}{c}{$ 1 $} 
                         & $ 1 $
                         & $-1 $\hphantom{$-$}
                         & $-1 $\hphantom{$-$}
\\
                         &
                         & $\mathrm{B}_{1}$
                         & \multicolumn{2}{c}{$ 1 $\hphantom{$-$}}
                         & \multicolumn{2}{c}{$-1 $\hphantom{$-$}} 
                         & $ 1 $
                         & $ 1 $
                         & $-1 $\hphantom{$-$}
\\
                         &
                         & $\mathrm{B}_{2}$
                         & \multicolumn{2}{c}{$ 1 $\hphantom{$-$}}
                         & \multicolumn{2}{c}{$-1 $\hphantom{$-$}} 
                         & $ 1 $
                         & $-1 $\hphantom{$-$}
                         & $ 1 $
\\
                         &
                         & $\mathrm{E}$
                         & \multicolumn{2}{c}{$ 2 $\hphantom{$-$}}
                         & \multicolumn{2}{c}{$ 0 $} 
                         & $-2 $\hphantom{$-$}
                         & $ 0 $
                         & $ 0 $
\\
\cline{3-10}
                         & \rule{0mm}{3mm}
                         & $\mathrm{E}_{\frac{1}{2}}$
                         & $ 2 $\hspace{-1.5mm}
                         & \hspace{-1.5mm}$-2 $\hphantom{$-$}
                         & \hspace{-1mm}$ \sqrt{2} $\hspace{-2mm}
                         & \hspace{-2mm}$-\sqrt{2} $\hphantom{$-$}\hspace{-1mm}
                         & $ 0 $
                         & $ 0 $
                         & $ 0 $
\\
                         & 
                         & $\mathrm{E}_{\frac{3}{2}}$
                         & $ 2 $\hspace{-1.5mm}
                         & \hspace{-1.5mm}$-2 $\hphantom{$-$}
                         & \hspace{-1mm}$-\sqrt{2} $\hphantom{$-$}\hspace{-2mm}
                         & \hspace{-2mm}$ \sqrt{2} $\hspace{-1mm}
                         & $ 0 $
                         & $ 0 $
                         & $ 0 $
\\[1.3mm]
\cline{3-10}
\\[-2.8mm]
\cline{3-10}
\end{tabular}
\label{T:characterC4vtilde}
\end{table}
\vspace{-15mm}
\begin{table}[htb]
\caption{Irreducible representations of the double group $\widetilde{\mathbf{C}_{2\mathrm{v}}}$
         and their characters.}
\begin{tabular}{cccccccccccccccc}
\cline{3-8}\\[-2.8mm]\cline{3-8}
                         &
                         & \rule{0mm}{6mm}
                         & \raisebox{1.5mm}[0mm][0mm]{$\bigl\{\overline{{E}}\bigr\}$}
                           \hspace{-1.0mm}
                         & \hspace{-1.0mm}
                           \raisebox{1.5mm}[0mm][0mm]{$\bigl\{\underline{{E}}\bigr\}$}
                         & \hspace{-2mm}
                           \shortstack{
                           $\bigl\{\overline{{C}_{2}},$ \\[-1mm]
                           $\,\,\,\,\,\underline{{C}_{2}}\,\bigr\}$}
                           \hspace{-2mm}
                         & \hspace{-2mm}
                           \shortstack{
                           $\bigl\{\overline{\sigma_{b}},$ \\[-1mm]
                           $\,\,\,\,\,\underline{\sigma_{b}}\,\bigr\}$}
                         & \hspace{-2mm}
                           \shortstack{
                           $\bigl\{\overline{\sigma_{a}},$ \\[-1mm]
                           $\,\,\,\,\,\underline{\sigma_{a}}\,\bigr\}$}
\\[1mm]
\cline{3-8}
\hspace{-2mm}
\raisebox
{-7.5mm}
[0mm][0mm]
{\smash{
 $\widetilde{\mathbf{C}_{2\mathrm{v}}} 
 \!
 \left\{
 \rule{0mm}{10.5mm}
 \right.$}}
\hspace{-5mm}            & \hspace{-2mm}
                           \raisebox
                           {-5mm}
                           [0mm][0mm]
                           {\smash{
                            $\mathbf{C}_{2\mathrm{v}}\!
                             \left\{
                             \rule{0mm}{8mm}
                             \right.$}}
                           \hspace{-5mm}
                         & $\mathrm{A}_{1}$
                         & \multicolumn{2}{c}{$ 1 $\hphantom{$-$}}
                         & $ 1 $
                         & \hphantom{$-$}$ 1 $\hphantom{$-$}
                         & \hphantom{$-$}$1$\hphantom{$-$}
\\
                         &
                         & $\mathrm{A}_{2}$
                         & \multicolumn{2}{c}{$ 1 $\hphantom{$-$}}
                         & $ 1 $
                         & $-1 $\hphantom{$-$}
                         & $-1 $\hphantom{$-$}
\\
                         &
                         & $\mathrm{B}_{1}$
                         & \multicolumn{2}{c}{$ 1 $\hphantom{$-$}}
                         & $-1 $\hphantom{$-$}
                         & $ 1 $
                         & $-1 $\hphantom{$-$}
\\
                         &
                         & $\mathrm{B}_{2}$
                         & \multicolumn{2}{c}{$ 1 $\hphantom{$-$}}
                         & $-1 $\hphantom{$-$}
                         & $-1 $\hphantom{$-$}
                         & $ 1 $
\\
\cline{3-8}
                         & 
                         & $\mathrm{E}_{\frac{1}{2}}$
                         & $ 2 $\hspace{-1.0mm}
                         & \hspace{-1.0mm}$-2 $\hphantom{$-$}
                         & $ 0 $
                         & $ 0 $
                         & $ 0 $
\\[1.3mm]
\cline{3-8}
\\[-2.8mm]
\cline{3-8}
\end{tabular}
\label{T:characterC2vtilde}
\end{table}
\vspace{-15mm}
\begin{table}[htb]
\caption{Irreducible representations of the double group $\widetilde{\mathbf{C}_{2}}$
         and their characters.}
\begin{tabular}{ccclcccccccccccccccccccccccccccccccc}
\cline{3-13}
\\[-2.8mm]
\cline{3-13}
     &                &
     & \rule{0mm}{3mm}
     & \multicolumn{2}{c}{$\{\overline{{E}}\}$}
     & \multicolumn{2}{c}{$\quad\{\underline{{E}}\}$}
     & 
     & \multicolumn{2}{c}{$\{\overline{C_{2}}\}$}
     & \multicolumn{2}{c}{$\quad\{\underline{C_{2}}\}$}
\\[0.5mm]
\cline{3-13}
\raisebox
{-7.5mm}
[0mm][0mm]
{\smash{
 $\widetilde{\mathbf{C}_{2}}
  \!
  \left\{
  \rule{0mm}{10.5mm}
  \right.$}}
\hspace{-6mm}
                         & \raisebox
                           {-1.7mm}
                           [0mm][0mm]
                           {\smash{
                            $\mathbf{C}_{2}\!
                             \left\{
                             \rule{0mm}{3.8mm}
                             \right.$}}
                           \hspace{-6mm}
                         & $ \mathrm{A} $
                         &
                         & \multicolumn{4}{c}{\hspace{-4mm}$ 1 $}
                         &
                         & \multicolumn{4}{c}{\hspace{-4mm}$ 1 $}
                         &
\\
                         &
                         & $ \mathrm{B} $
                         &
                         & \multicolumn{4}{c}{\hspace{-4mm}$ 1 $}
                         &
                         & \multicolumn{4}{c}{\hspace{-5.5mm}$-1 $\vphantom{$-$}}
                         & 
\\\cline{3-13}
                         & \rule{0mm}{3.5mm}
                         & \raisebox{-3mm}
                           [0mm][0mm]
                           {$ \mathrm{E}_{\frac{1}{2}} $}
                         & \hspace{-6mm}
                           \raisebox
                           {-3mm}
                           [0mm][0mm]
                           {\smash{
                            $\left\{
                            \rule{0mm}{6mm}
                            \right.$}}
                           \hspace{-1.5mm}
                           $\mathrm{E}_{\frac{1}{2}}^{(1)}$
                         & \multicolumn{2}{l}
                           {\raisebox{-3mm}
                           [0mm][0mm]
                           {$ 2 $}
                           \hspace{-2mm}
                           \raisebox
                           {-3mm}
                           [0mm][0mm]
                           {\smash{
                            $\left\{
                            \rule{0mm}{6mm}
                            \right.$}}
                           \hspace{-2mm}
                           $ 1 $
                           }
                         & \multicolumn{2}{c}
                           {\raisebox{-3mm}
                           [0mm][0mm]
                           {$-2 $}
                           \hspace{-2mm}
                           \raisebox
                           {-3mm}
                           [0mm][0mm]
                           {\smash{
                            $\left\{
                            \rule{0mm}{6mm}
                            \right.$}}
                           \hspace{-2mm}
                           $-1 $\hspace{-3mm}
                           }
                         &
                         & \multicolumn{2}{c}
                           {\hspace{-1mm}\raisebox{-3mm}
                           [0mm][0mm]
                           {$ 0 $}
                           \hspace{-2mm}
                           \raisebox
                           {-3mm}
                           [0mm][0mm]
                           {\smash{
                            $\left\{
                            \rule{0mm}{6mm}
                            \right.$}}
                           \hspace{-2mm}
                           $-{i} $\hspace{-4mm}
                           }
                         & \multicolumn{2}{c}
                           {\raisebox{-3mm}
                           [0mm][0mm]
                           {$ 0 $}
                           \hspace{-2mm}
                           \raisebox
                           {-3mm}
                           [0mm][0mm]
                           {\smash{
                            $\left\{
                            \rule{0mm}{6mm}
                            \right.$}}
                           \hspace{-2mm}
                           $ {i} $\hspace{-4mm}
                           }
\\[2mm]
                         &
                         &
                         & $\hspace{-2.2mm}\mathrm{E}_{\frac{1}{2}}^{(2)} $
                         &
                         & \!\!\!$ 1 $
                         &
                         & \,$-1 $\hspace{-3mm}
                         & 
                         &
                         & $\!\!\!\!\!{i}$\hspace{-3mm}
                         &
                         & \,\,\,$\quad-{i}$
\\[2mm]
\cline{3-13}
\\[-2.8mm]
\cline{3-13}
\end{tabular}
\label{T:characterC2tilde}
\end{table}
\begin{figure*}[b]
\vspace*{-3mm}
\centering
\includegraphics[width=177mm]{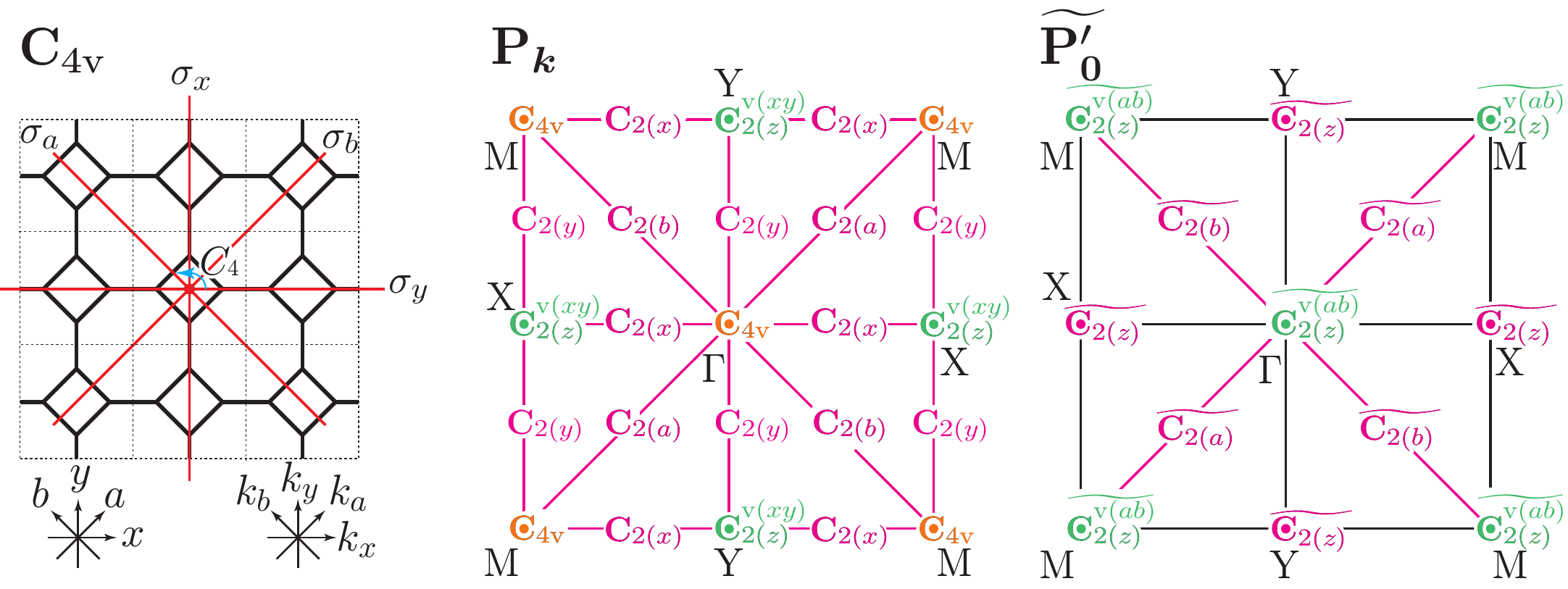}
\vspace*{-5mm}
\caption{The diamond-square lattice of $\mathbf{C}_{4\mathrm{v}}\equiv\mathbf{P}_{\mathrm{org}}$
         point symmetry with primitive cells encircled by dotted lines.
         $\bm{k}$-point symmetry groups (isotropy groups of $\bm{k}$) $\mathbf{P}_{\bm{k}}$
         at high symmetry points $\bm{k}$ of the diamond-square reciprocal lattice,
         each consisting of symmetry operations $P_{\bm{k}}$ such that
         $P_{\bm{k}}\bm{k}=\bm{k}+\bm{K}\cong\bm{k}$ with
         $\bm{K}$ being $\bm{0}$ or a reciprocal lattice vector.
         $\mathbb{Z}_2$-gauged $\bm{k}$-point symmetry groups
         $\widetilde{\mathbf{P}_{\bm{k}}'}$ under the Fourier transformation (\ref{E:FT}),
         i.e. $\widetilde{\mathbf{P}_{\bm{0}}'}$, at high symmetry points $\bm{k}_\kappa$ of
         the diamond-square reciprocal lattice,
         each keeping the $\bm{k}_\kappa$ block $\mathcal{H}_{\bm{k}_\kappa}$ of
         the Fourier-transformed gauge-ground Majorana Hamiltonian
         (\ref{E:HMajoranaGSk}) invariant.
         Note that
         $\mathbf{P}_{\bm{k}}'\subseteq\mathbf{P}_{\bm{k}}\subseteq\mathbf{P}_{\mathrm{org}}$.}
\label{F:prSGofk}
\end{figure*}

\section*{S3. Direct-Product Representations of Double Groups for
          the Gauged Diamond-Square Lattice}
\label{S:ChTDPrepstildeP}

   Inelastic visible-light scatterings within the Loudon-Fleury scheme
\cite{F514,S1068,S365,K187201}
are mediated by ``momentum-locked" spinon geminate excitations
$\alpha_{\bm{k}_\kappa:\lambda}^\dagger\alpha_{-\bm{k}_\kappa:\lambda'}^\dagger$.
In order to analyze the polarized Raman spectra of the gauged diamond-square lattice
Fig. \ref{F:prSG}(d), we formulate direct-product representations of gauged $\bm{k}$-point
symmetry groups
$\widetilde{\mathbf{P}_{\bm{k}}'}
\subseteq\widetilde{\mathbf{P}}
=\widetilde{\mathbf{C}_{4\mathrm{v}}}$ at high symmetry points of the diamond-square reciprocal
lattice.
When we need to specify the principal axis $\tau$ for $n$-fold rotations and/or
the normal vector $\sigma$ for mirror operations,
we replace the usual Sch\"onflies notation $\mathbf{C}_{n(\tau)\mathrm{v}(\sigma)}$
by $\mathbf{C}_{n(\tau)}^{\mathrm{v}(\sigma)}$ for the sake of saving space.

   The isotropy group of $\bm{k}$ in the first Brillouin zone $\mathbf{P}_{\bm{k}}$ consists of
symmetry operations $P_{\bm{k}}$ such that
\begin{align}
   P_{\bm{k}}\bm{k}
  =\bm{k}+\bm{K}
  \cong
   \bm{k}
\end{align}
with $\bm{K}$ being $\bm{0}$ or a reciprocal lattice vector.
The  $\mathbb{Z}_2$-gauged $\bm{k}$-point symmetry group $\widetilde{\mathbf{P}_{\bm{k}}'}$
keeps the $\bm{k}_\kappa$ block $\mathcal{H}_{\bm{k}_\kappa}$ of the gauge-ground Majorana
Hamiltonian (\ref{E:HMajoranaGSr}) invariant.
Under the present Fourier transformation (\ref{E:FT}),
$\widetilde{\mathbf{P}_{\bm{k}}'}$ should consist of \textit{primitive-translation-invariant}
gauged point symmetry operations, i.e., $\widetilde{\mathbf{P}_{\bm{k}}'}$ may be written as
$\widetilde{\mathbf{P}_{\bm{0}}'}$ with
$\mathbf{P}_{\bm{0}}'\subseteq\mathbf{P}_{\bm{0}}$.
$\mathbf{P}_{\bm{0}}$ equals the full point symmetry group of the background lattice
$\mathbf{P}_{\mathrm{org}}$ and reads $\mathbf{C}_{4\mathrm{v}}$ for the diamond-square lattice.
Figure \ref{F:prSG}(d) shows that the gauge transformations
$\varLambda(C_4)$, $\varLambda(C_4^{-1})$, $\varLambda(\sigma_x)$, and $\varLambda(\sigma_y)$
recover the initial ground gauge configuration but break the primitive translation symmetry.
Such gauged point symmetry operations keep none of the original $\bm{k}_\kappa$ blocks
$\mathcal{H}_{\bm{k}_\kappa}$ $(\kappa=1,\cdots,N^2)$ invariant and
demand that the Fourier transformation (\ref{E:FT}) be modified.
Thus, every $\mathbf{P}_{\bm{0}}'$ is limited to a proper subset of $\mathbf{P}_{\bm{0}}$,
$\mathbf{P}_{\bm{0}}'\subset\mathbf{P}_{\bm{0}}$,
for the diamond-square lattice.
We show in Fig. \ref{F:prSGofk} how $\mathbf{P}_{\bm{k}}$ and $\widetilde{\mathbf{P}_{\bm{0}}'}$
read at high symmetry points of the diamond-square reciprocal lattice.
$\widetilde{\mathbf{P}_{\bm{0}}'}$ becomes $\widetilde{\mathbf{C}_{2\mathrm{v}}}$ even at
the highest symmetry points $\Gamma$ and $\mathrm{M}$.
The thus-defined projective symmetry group of the $\mathbb{Z}_2$-gauged diamond-square lattice
reads
$
   \mathbf{L}\wedge\widetilde{\mathbf{C}_{2\mathrm{v}}}
$
with
$\mathbf{L}=\{m\bm{a}|m\in\mathbb{Z}\}\times\{n\bm{b}|n\in\mathbb{Z}\}$.

   Each momentum-locked spinon geminate excitation
$\alpha_{\bm{k}_\kappa:\lambda}^\dagger\alpha_{-\bm{k}_\kappa:\lambda'}^\dagger$
consists of double-valued irreducible representations of the gauged $\bm{k}$-point symmetry group
$\widetilde{\mathbf{P}_{\bm{0}}'}$ at points $\pm\bm{k}_\kappa$ and have a direct-product
representation,
$\widetilde{\varXi}_i\otimes\widetilde{\varXi}_{i'}$
$\bigl(
  i,i'=n_{\mathcal{C}}^{\mathbf{P}_{\bm{0}}'}+1,\cdots,
       n_{\mathcal{C}}^{\widetilde{\mathbf{P}_{\bm{0}}'}}
 \bigr)$.
Since its characters read
\begin{align}
   \chi_{\widetilde{\varXi}_i\otimes\widetilde{\varXi}_{i'}}
   ^{\widetilde{\mathbf{P}}_{\bm{0}}'}(\overline{P_{\bm{0}}'})
   &
  =\chi_{\widetilde{\varXi}_i}^{\widetilde{\mathbf{P}}_{\bm{0}}'}(\overline{P_{\bm{0}}'})
   \chi_{\widetilde{\varXi}_{i'}}^{\widetilde{\mathbf{P}}_{\bm{0}}'}(\overline{P_{\bm{0}}'})
   \nonumber \\
   &
  =\chi_{\widetilde{\varXi}_i}^{\widetilde{\mathbf{P}}_{\bm{0}}'}(\underline{P_{\bm{0}}'})
   \chi_{\widetilde{\varXi}_{i'}}^{\widetilde{\mathbf{P}}_{\bm{0}}'}(\underline{P_{\bm{0}}'})
  =\chi_{\widetilde{\varXi}_i\otimes\widetilde{\varXi}_{i'}}
   ^{\widetilde{\mathbf{P}}_{\bm{0}}'}(\underline{P_{\bm{0}}'}),
   \label{E:character(DPRij)}
\end{align}
it results in single-valued irreducible representations of $\widetilde{\mathbf{P}_{\bm{0}}'}$,
i.e. those of the corresponding point symmetry group $\mathbf{P}_{\bm{0}}'$,
\vspace*{-1mm}
\begin{align}
   \!\!\!\!
   \widetilde{\varXi}_{i}\otimes\widetilde{\varXi}_{i'}
   &
  =\mathop{\bigoplus}_{j=1}^{n_{\mathcal{C}}^{\widetilde{\mathbf{P}}_{\bm{0}}'}}
   \widetilde{\varXi}_j
   \sum_{q=1}^{n_{\mathcal{C}}^{\widetilde{\mathbf{P}}_{\bm{0}}'}}
      \frac{h_q}{g^{\widetilde{\mathbf{P}}_{\bm{0}}'}}
   \chi_{\widetilde{\varXi}_j}^{\widetilde{\mathbf{P}}_{\bm{0}}'}(\mathcal{C}_{q})^*
   \chi_{\widetilde{\varXi}_i\otimes\widetilde{\varXi}_{i'}}^{\widetilde{\mathbf{P}}_{\bm{0}}'}
   (\mathcal{C}_{q})
   \nonumber \\
   &
  =\mathop{\bigoplus}_{j=1}^{n_{\mathcal{C}}^{\mathbf{P}_{\bm{0}}'}}\varXi_j
   \sum_{q=1}^{n_{\mathcal{C}}^{\widetilde{\mathbf{P}}_{\bm{0}}'}}
      \frac{h_q}{g^{\widetilde{\mathbf{P}}_{\bm{0}}'}}
   \chi_{\varXi_j}^{\mathbf{P}_{\bm{0}}'}(\mathcal{C}_{q})^*
   \chi_{\widetilde{\varXi}_i\otimes\widetilde{\varXi}_{i'}}^{\widetilde{\mathbf{P}}_{\bm{0}}'}
   (\mathcal{C}_{q}).
   \label{E:DPR}
   \\[-8mm]\nonumber
\end{align}
When we make a direct product of the same double-valued irreducible representations of
$\widetilde{\mathbf{P}_{\bm{0}}'}$, it reads a direct sum of symmetric and antisymmetric
single-valued irreducible representations of $\mathbf{P}_{\bm{0}}'$,
\begin{align}
   &
   \widetilde{\varXi}_{i}\otimes\widetilde{\varXi}_{i}
  =[\widetilde{\varXi}_{i}\otimes\widetilde{\varXi}_{i}]
   \oplus
   \{\widetilde{\varXi}_{i}\otimes\widetilde{\varXi}_{i}\};
   \label{E:DPRii=symDPRii+antisymDPRii}
   \\
   &
   [\widetilde{\varXi}_i\otimes\widetilde{\varXi}_i]
  =\mathop{\bigoplus}_{j=1}^{n_{\mathcal{C}}^{\mathbf{P}_{\bm{0}}'}}[\varXi_{j}]
   \sum_{q=1}^{n_{\mathcal{C}}^{\widetilde{\mathbf{P}_{\bm{0}}'}}}
      \frac{h_{q}}{g^{\widetilde{\mathbf{P}_{\bm{0}}'}}}
   \chi_{\varXi_{j}}^{\mathbf{P}_{\bm{0}}'}(\mathcal{C}_{q})^*
   \chi_{[\widetilde{\varXi}_{i}\otimes\widetilde{\varXi}_{i}]}
   ^{\widetilde{\mathbf{P}_{\bm{0}}'}}
   (\mathcal{C}_{q}),
   \label{E:symDPRii}
   \\
   &
   \{\widetilde{\varXi}_i\otimes\widetilde{\varXi}_i\}
  =\mathop{\bigoplus}_{j=1}^{n_{\mathcal{C}}^{\mathbf{P}_{\bm{0}}'}}\{\varXi_{j}\}
   \sum_{q=1}^{n_{\mathcal{C}}^{\widetilde{\mathbf{P}_{\bm{0}}'}}}
      \frac{h_{q}}{g^{\widetilde{\mathbf{P}_{\bm{0}}'}}}
   \chi_{\varXi_{j}}^{\mathbf{P}_{\bm{0}}'}(\mathcal{C}_{q})^*
   \chi_{\{\widetilde{\varXi}_{i}\otimes\widetilde{\varXi}_{i}\}}
   ^{\widetilde{\mathbf{P}_{\bm{0}}'}}
   (\mathcal{C}_{q}),
   \label{E:antisymDPRii}
   \\
   &
   \chi_{[\widetilde{\varXi}_{i}\otimes\widetilde{\varXi}_{i}]}^{\widetilde{\mathbf{P_{\bm{0}}}}}
   (\widetilde{P_{\bm{0}}'})
  =\frac{1}{2}
   \left[
      \chi_{\widetilde{\varXi}_{i}}^{\widetilde{\mathbf{P}_{\bm{0}}'}}(\widetilde{P_{\bm{0}}'})^{2}
     +\chi_{\widetilde{\varXi}_{i}}^{\widetilde{\mathbf{P}_{\bm{0}}'}}(\widetilde{P_{\bm{0}}'}^{2})
   \right],
   \label{E:character(symDPRii)}
   \\
   &
   \chi_{\{\varXi_{i}\otimes\varXi_{i}\}}^{\widetilde{\mathbf{P}_{\bm{0}}'}}
   (\widetilde{P_{\bm{0}}'})
  =\frac{1}{2}
   \left[
     \chi_{\widetilde{\varXi}_{i}}^{\widetilde{\mathbf{P}_{\bm{0}}'}}(\widetilde{P_{\bm{0}}'})^{2}
    -\chi_{\widetilde{\varXi}_{i}}^{\widetilde{\mathbf{P}_{\bm{0}}'}}(\widetilde{P_{\bm{0}}'}^{2})
   \right].
   \label{E:character(antisymDPRii)}
\end{align}
\clearpage

   We can obtain characters of any direct-product representation using
Eqs. (\ref{E:character(symDPRii)}) and (\ref{E:character(antisymDPRii)}) as well as
(\ref{E:character(DPRij)}),
which of our interest are listed in
Tables \ref{T:characterDPRC2vtilde} and \ref{T:characterDPRC2tilde}.
Direct-product representations for geminate excitations of different Majorana spinon eigenmodes
are not necessarily made of different irreducible representations but may be made of the same ones.
Those made of different irreducible representations can be decomposed into irreducible
representations by Eq. (\ref{E:DPR}), while those made of the same ones by
Eqs. (\ref{E:symDPRii}) and (\ref{E:antisymDPRii}).
Direct-product representations for geminate excitations of degenerate Majorana spinon eigenmodes
are also the latter case.
The thus-obtained decompositions into irreducible representations are all listed in
Table \ref{T:DPRintoIrrep}.

   Any symmetry species $\varXi_l$ of the $\widetilde{\mathbf{C}_{4\mathrm{v}}}$-gauged-lattice
Raman spectrum belongs to $\mathbf{C}_{4\mathrm{v}}$, while spinon geminate excitations
yield one or more irreducible representations $\bigoplus_j\varXi_j$ of its subgroup
$\mathbf{P}_{\bm{0}}'\subset\mathbf{P}_{\bm{0}}=\mathbf{C}_{4\mathrm{v}}$.
When $\mathbf{P}_{\bm{0}}'\subset\mathbf{C}_{4\mathrm{v}}$,
every irreducible representation $\varXi_j$ of $\mathbf{P}_{\bm{0}}'$ is compatible
with one or more irreducible representations $\varXi_l$ of $\mathbf{C}_{4\mathrm{v}}$,
in other words,
every irreducible representation $\varXi_l$ of $\mathbf{C}_{4\mathrm{v}}$ reads
a direct sum of one or more irreducible representations
$\varXi_j\in\mathbf{P}_{\bm{0}}'$.
When we denote the representation $\varXi_l$ of $\mathbf{C}_{4\mathrm{v}}$ within its subgroup
$\mathbf{P}_{\bm{0}}'$ by $\varXi_l\downarrow\mathbf{P}_{\bm{0}}'$,
the \textit{compatibility relation} reads
\begin{align}
   \varXi_l\downarrow\mathbf{P}_{\bm{0}}'
  =\mathop{\bigoplus}_{j=1}^{n_{\mathcal{C}}^{\mathbf{P}_{\bm{0}}'}}\varXi_j
   \sum_{q=1}^{n_{\mathcal{C}}^{\mathbf{P}_{\bm{0}}'}}
      \frac{h_q}{g^{\mathbf{P}_{\bm{0}}'}}
   \chi_{\varXi_j}^{\mathbf{P}_{\bm{0}}'}(\mathcal{C}_{q})^*
   \chi_{\varXi_l\downarrow\mathbf{P}_{\bm{0}}'}^{\mathbf{P}_{\bm{0}}'}
   (\mathcal{C}_{q})
   \label{E:IrrepDecompose}
\end{align}
with
$
   \chi_{\varXi_{l}\downarrow \mathbf{P}_{\bm{0}}'}^{\mathbf{P}_{\bm{0}}'}(P_{\bm{0}}')
  =\chi_{\varXi_{l}}^{\mathbf{C}_{4\mathrm{v}}}(P_{\bm{0}}')
$.
Compatibility relations between point symmetry groups generally depend on their principal
rotation axes and mirror normal vectors.
The classes of those in question are given by
\vspace*{-1mm}
\begin{align}
   \mathbf{C}_{4\mathrm{v}}
   &
  :\{E\},\,\{2C_{4(z)}\},\,\{C_{2(z)}\},\,\{\sigma_{x},\sigma_{y}\},
   \{\sigma_{a},\sigma_{b}\},
   \nonumber
   \\
   \mathbf{C}_{2(z)}^{\mathrm{v}(ab)}
   &
  :\{E\},\,\{C_{2(z)}\},\,\{\sigma_{a}\},\{\sigma_{b}\},
   \nonumber
   \\
   \mathbf{C}_{2(\tau)}
   &
  :\{E\},\,\{C_{2(\tau)}\}\quad(\tau = z,x,y,a,b).
   \nonumber
   \\[-8mm]\nonumber
\end{align}
Let us find compatible representations of $\mathbf{C}_{2(z)}^{\mathrm{v}(ab)}$ and
$\mathbf{C}_{2(z)}$ for $\mathrm{E}\in\mathbf{C}_{4\mathrm{v}}$.
Note that
$g^{\mathbf{C}_{2(z)}}=2$, $n_{\mathcal{C}}^{\mathbf{C}_{2(z)}}=2$, and $h_1=h_2=1$
for $\mathbf{C}_{2(z)}$, while
$g^{\mathbf{C}_{2(z)}^{\mathrm{v}(ab)}}=4$,
$n_{\mathcal{C}}^{\mathbf{C}_{2(z)}^{\mathrm{v}(ab)}}=4$, and $h_1=\cdots=h_4=1$
for $\mathbf{C}_{2(z)}^{\mathrm{v}(ab)}$.
Table \ref{T:characterC4vtilde} shows that
\begin{alignat}{7}
   &
   \chi_{\mathrm{E}\downarrow\mathbf{C}_{2(z)}}^{\mathbf{C}_{2(z)}}(E)
   &&
  =&&\chi_{\mathrm{E}\downarrow\mathbf{C}_{2(z)}^{\mathrm{v}(ab)}}
         ^{\mathbf{C}_{2(z)}^{\mathrm{v}(ab)}}(E)
   &&
  =\chi_{\mathrm{E}}^{\mathbf{C}_{4\mathrm{v}}}(E)
   &&
  =& 2,
   \nonumber \\
   &
   \chi_{\mathrm{E}\downarrow\mathbf{C}_{2(z)}}^{\mathbf{C}_{2(z)}}(C_{2(z)})
   &&
  =&&\chi_{\mathrm{E}\downarrow\mathbf{C}_{2(z)}^{\mathrm{v}(ab)}}
         ^{\mathbf{C}_{2(z)}^{\mathrm{v}(ab)}}(C_{2(z)})
   &&
  =\chi_{\mathrm{E}}^{\mathbf{C}_{4\mathrm{v}}}(C_{2(z)})
   &&
  =&-2,
   \nonumber \\
   &
   &&
   &&
   \chi_{\mathrm{E}\downarrow\mathbf{C}_{2(z)}^{\mathrm{v}(ab)}}
       ^{\mathbf{C}_{2(z)}^{\mathrm{v}(ab)}}(\sigma_a)
   &&
  =\chi_{\mathrm{E}}^{\mathbf{C}_{4\mathrm{v}}}(\sigma_a)
   &&
  =& 0,
   \nonumber \\
   &
   &&
   &&
   \chi_{\mathrm{E}\downarrow\mathbf{C}_{2(z)}^{\mathrm{v}(ab)}}
       ^{\mathbf{C}_{2(z)}^{\mathrm{v}(ab)}}(\sigma_b)
   &&
  =\chi_{\mathrm{E}}^{\mathbf{C}_{4\mathrm{v}}}(\sigma_b)
   &&
  =& 0,
\end{alignat}
while all the characters for $\mathbf{C}_{2(z)}^{\mathrm{v}(ab)}$ and $\mathbf{C}_{2(z)}$
are available from Tables \ref{T:characterC2vtilde} and \ref{T:characterC2tilde}.
Then Eq. (\ref{E:IrrepDecompose}) yields
\begin{align}
   &
   \mathrm{E}\downarrow\mathbf{C}_{2(z)}^{\mathrm{v}(ab)}
  =\sum_{j=1}^4\sum_{q=1}^4
    \frac{\varXi_j}{4}
    \chi_{\varXi_j}^{\mathbf{C}_{2(z)}^{\mathrm{v}(ab)}}(\mathcal{C}_{q})
    \chi_{\mathrm{E}}^{\mathbf{C}_{4\mathrm{v}}}(\mathcal{C}_{q})
  =\mathrm{B}_1\oplus\mathrm{B}_2,
   \nonumber \\
   &
   \mathrm{E}\downarrow\mathbf{C}_{2(z)}
  =\sum_{j=1}^2\sum_{q=1}^2
    \frac{\varXi_j}{2}
    \chi_{\varXi_j}^{\mathbf{C}_{2(z)}}(\mathcal{C}_{q})
    \chi_{\mathrm{E}}^{\mathbf{C}_{4\mathrm{v}}}(\mathcal{C}_{q})
  =2\mathrm{B}.
\end{align}
We summarize in Table \ref{T:CompatibilityRelationC4vC2vC2}
the thus-obtained compatibility relations between $\mathbf{C}_{4\mathrm{v}}$ and
its subgroups $\mathbf{C}_{2\mathrm{v}}$ and $\mathbf{C}_2$.
\vspace*{-4mm}

\begin{table}[htb]
\caption{Direct-product representations made of double-valued irreducible representations of
         the double group $\widetilde{\mathbf{C}_{2\mathrm{v}}}$ and their characters.}
\begin{tabular}{rclccccccccccccccccc}
\hline\hline
\raisebox{1.5mm}[0mm][0mm]{$\widetilde{\varXi}_{i}$} & \hspace{-4.5mm}
\raisebox{1.5mm}[0mm][0mm]{$\otimes$}                & \hspace{-4.5mm}
\raisebox{1.5mm}[0mm][0mm]{$\widetilde{\varXi}_{j}$} 
                         & \rule{0mm}{6.5mm}
                           \raisebox{1.5mm}[0mm][0mm]{$\{\overline{{E}}\}$}\hspace{-1.0mm}
                         & \hspace{-1.0mm}\raisebox{1.5mm}[0mm][0mm]{$\{\underline{{E}}\}$}
                         & \hspace{-2mm}
                           \shortstack{
                           $\{\overline{{C}_{2}},$ \\
                           $\,\,\,\,\,\underline{{C}_{2}}\,\}$}
                         & \hspace{-2mm}
                           \shortstack{
                           $\{\overline{\sigma_{b}},$ \\
                           $\,\,\,\,\,\underline{\sigma_{b}}\,\}$}
                         & \hspace{-2mm}
                           \shortstack{
                           $\{\overline{\sigma_{a}},$ \\
                           $\,\,\,\,\,\underline{\sigma_{a}}\,\}$}
\\[1mm]
\hline
$[\mathrm{E}_{\frac{1}{2}}$    & \hspace{-4.5mm}
$\otimes$                      & \hspace{-4.5mm}
$\mathrm{E}_{\frac{1}{2}}]$    & \multicolumn{2}{c}{$ 3 $}
                               & $-1 $\hphantom{$-$}
                               & $-1 $\hphantom{$-$}
                               & $-1 $\hphantom{$-$}
\\
$\{\mathrm{E}_{\frac{1}{2}}$   & \hspace{-4.5mm}
$\otimes$                      & \hspace{-4.5mm}
$\mathrm{E}_{\frac{1}{2}}\}$   & \multicolumn{2}{c}{$ 1 $}
                               & \vphantom{$-$}$ 1 $
                               & \vphantom{$-$}$ 1 $
                               & \vphantom{$-$}$ 1 $
\\[1.5mm]
\hline\hline
\end{tabular}
\label{T:characterDPRC2vtilde}
\centering
\caption{Direct-product representations made of double-valued irreducible representations of
         the double group $\widetilde{\mathbf{C}_{2}}$ and their characters.}
\begin{tabular}{rclccccccccccccccccc}
\hline\hline
\rule{0mm}{3mm}
\raisebox{0mm}[0mm][0mm]{$\widetilde{\varXi}_{i}$} & \hspace{-4.5mm}
\raisebox{0mm}[0mm][0mm]{$\otimes$}                & \hspace{-4.5mm}
\raisebox{0mm}[0mm][0mm]{$\widetilde{\varXi}_{j}$} 
                         & \raisebox{0mm}[0mm][0mm]{$\{\overline{{E}}\}$}\hspace{-1.0mm}
                         & \hspace{-1.0mm}\raisebox{0mm}[0mm][0mm]{$\{\underline{{E}}\}$}
                         & \raisebox{0mm}[0mm][0mm]{$\{\overline{{C}_{2}}\}$}\hspace{-1.0mm}
                         & \hspace{-1.0mm}\raisebox{0mm}[0mm][0mm]{$\{\underline{{C}_{2}}\}$}
\\[0.5mm]
\hline
\rule{0mm}{3mm}
$[\mathrm{E}_{\frac{1}{2}}^{(1)}$   & \hspace{-4.5mm}
$\otimes$                           & \hspace{-4.5mm}
$\mathrm{E}_{\frac{1}{2}}^{(1)}]$   & \multicolumn{2}{c}{$ 1 $}
                                    & \multicolumn{2}{c}{$-1 $\hphantom{$-$}}
\\[2mm]
$\mathrm{E}_{\frac{1}{2}}^{(1)}$    & \hspace{-4.5mm}
$\otimes$                           & \hspace{-4.5mm}
$\mathrm{E}_{\frac{1}{2}}^{(2)}$    & \multicolumn{2}{c}{$ 1 $}
                                    & \multicolumn{2}{c}{$ 1 $}
\\[2mm]
$[\mathrm{E}_{\frac{1}{2}}^{(2)}$   & \hspace{-4.5mm}
$\otimes$                           & \hspace{-4.5mm}
$\mathrm{E}_{\frac{1}{2}}^{(2)}]$   & \multicolumn{2}{c}{$ 1 $}
                                    & \multicolumn{2}{c}{$-1 $\hphantom{$-$}}
\\[2mm]
\hline\hline
\end{tabular}
\label{T:characterDPRC2tilde}
\centering
\centering
\caption{Direct-product representations made of double-valued irreducible representations
         $\widetilde{\varXi}_i\otimes\widetilde{\varXi}_{i'}$
         and their decompositions into single-valued irreducible representations
         $\widetilde{\varXi}_j$, which are underlined when they are relevant to Raman scattering,
         for gauged $\bm{k}$-point symmetry groups $\widetilde{\mathbf{P}_{\bm{k}}'}$.}
\vspace{-2mm}
\begin{tabular}{crcclrccc}
\hline \hline
\rule{0mm}{3mm}
$\widetilde{\mathbf{P}_{\bm{k}}'}$
                                       & \multicolumn{3}{c}
                                         {$\widetilde{\varXi}_i\otimes\widetilde{\varXi}_{i'}$}
                                       & $\bigoplus_j\widetilde{\varXi}_j
                                         =\bigoplus_j\varXi_j$
\\[0.5mm]\hline
\raisebox{-3.5mm}[0mm][0mm]{
$\left\{\rule{0mm}{5mm}\right.$
}
\hspace{-3mm}
\raisebox{-6mm}[0mm][0mm]{
\shortstack{
$\widetilde{\mathbf{C}_{2(x)}},
 \widetilde{\mathbf{C}_{2(y)}},$
\\
$\widetilde{\mathbf{C}_{2(a)}},
 \widetilde{\mathbf{C}_{2(b)}}$
}
}
\hspace{-3mm}
\raisebox{-3.5mm}[0mm][0mm]{
$\left.\rule{0mm}{5mm}\right\}$
}
\rule{0mm}{3.2mm}
                                       & \multicolumn{3}{c}
                                         {\,\,$\mathrm{E}_{\frac{1}{2}}^{(1)}
                                           \otimes
                                           \mathrm{E}_{\frac{1}{2}}^{(1)},\,\,
                                           \mathrm{E}_{\frac{1}{2}}^{(2)}
                                           \otimes
                                           \mathrm{E}_{\frac{1}{2}}^{(2)}
                                          $
                                         }
                                       & \,$\bigl[\underline{\mathrm{B}}\bigr]$
                                       \\[2mm]
                                       & \multicolumn{3}{c}
                                         {\,\,$\mathrm{E}_{\frac{1}{2}}^{(1)}
                                           \otimes
                                           \mathrm{E}_{\frac{1}{2}}^{(2)},\,\,
                                           \mathrm{E}_{\frac{1}{2}}^{(2)}
                                           \otimes
                                           \mathrm{E}_{\frac{1}{2}}^{(1)}
                                          $
                                         }
                                       & \,\ $\underline{\mathrm{A}}$
\\[2mm]\hline
\raisebox{-4mm}[0mm][0mm]{$\widetilde{\mathbf{C}_{2(z)}}$}\rule{0mm}{3.2mm}
                                       & \multicolumn{3}{c}
                                         {\,\,$\mathrm{E}_{\frac{1}{2}}^{(1)}
                                           \otimes
                                           \mathrm{E}_{\frac{1}{2}}^{(1)},\,\,
                                           \mathrm{E}_{\frac{1}{2}}^{(2)}
                                           \otimes
                                           \mathrm{E}_{\frac{1}{2}}^{(2)}
                                          $
                                         }
                                       & \,$\bigl[\mathrm{B}\bigr]$
                                       \\[2mm]
                                       & \multicolumn{3}{c}
                                         {\,\,$\mathrm{E}_{\frac{1}{2}}^{(1)}
                                           \otimes
                                           \mathrm{E}_{\frac{1}{2}}^{(2)},\,\,
                                           \mathrm{E}_{\frac{1}{2}}^{(2)}
                                           \otimes
                                           \mathrm{E}_{\frac{1}{2}}^{(1)}
                                          $
                                         }
                                       & \,\ $\underline{\mathrm{A}}$
\\[2mm]\hline
$\widetilde{\mathbf{C}_{2(z)}^{\mathrm{v}(ab)}}$\rule{0mm}{3.8mm}
                                       & \multicolumn{3}{c}
                                         {$\mathrm{E}_{\frac{1}{2}}
                                           \otimes
                                           \mathrm{E}_{\frac{1}{2}}
                                          $
                                         }
                                       & $\bigl\{\underline{\mathrm{A}_{1}}\bigr\}\oplus
                                          \bigl[\underline{\mathrm{A}_{2}}\bigr]\oplus
                                          \bigl[\mathrm{B}_{1}\bigr]\oplus
                                          \bigl[\mathrm{B}_{2}\bigr]$
\\[1.3mm]\hline\hline
\end{tabular}
\label{T:DPRintoIrrep}
\centering
\caption{Compatibility relations between irreducible representations of
         $\mathbf{C}_{4\mathrm{v}}$ and those of its subgroups $\mathbf{C}_{2\mathrm{v}}$ and
         $\mathbf{C}_2$.
         The $\mathbb{Z}_2$-gauged diamond-square lattice is invariant to
         $\widetilde{\mathbf{C}_{4\mathrm{v}}}\equiv\widetilde{\mathbf{P}}$
         [Fig. \ref{F:prSG}(d)],
         while each block $\mathcal{H}_{\bm{k}_\kappa}$ of the gauge-ground Majorana Hamiltonian
         (\ref{E:HMajoranaGSr}) belongs to its subgroup $\widetilde{\mathbf{P}_{\bm{k}}'}$,
         where $\mathbf{P}_{\bm{k}}'\subseteq\mathbf{P}_{\bm{k}}\subseteq\mathbf{P}$
         (Fig. \ref{F:prSGofk}).
         When we adopt the Fourier transformation (\ref{E:FT}),
         $\widetilde{\mathbf{P}_{\bm{k}}'}$ reads $\widetilde{\mathbf{P}_{\bm{0}}'}$ with
         $\mathbf{P}_{\bm{0}}=\mathbf{C}_{4\mathrm{v}}$.
         The Raman-active modes of $\mathbf{C}_{4\mathrm{v}}$ are underlined and their
         compatible symmetry species of $\mathbf{P}_{\bm{k}}'$ determine
         \textit{which type of spinon geminate excitations is relevant to which symmetry
         species of the Raman scattering intensities}.}
\begin{tabular}{cccccccc}\hline\hline
  \rule{0mm}{3mm}
  $\mathbf{C}_{4\mathrm{v}}$
  & $\mathrm{A}_{1}$                     & $\mathrm{A}_{2}$
  & $\underline{\mathrm{B}_{1}}$         & $\underline{\mathrm{B}_{2}}$
  & $\mathrm{E}$
  \\[0.8mm] \hline
  \rule{0mm}{3mm}
  $\mathbf{C}_{2(z)}^{\mathrm{v}(ab)}$
  & $\mathrm{A}_{1}$                     & $\mathrm{A}_{2}$
  & $\mathrm{A}_{2}$                     & $\mathrm{A}_{1}$
  & $\mathrm{B}_{1}\oplus\mathrm{B}_{2}$ 
  \\
  $\mathbf{C}_{2(z)}$
  & $\mathrm{A}$                         & $\mathrm{A}$
  & $\mathrm{A}$                         & $\mathrm{A}$
  & $2\mathrm{B}$
  \\
  $\mathbf{C}_{2(x)},\mathbf{C}_{2(y)}$
  & $\mathrm{A}$                         & $\mathrm{B}$
  & $\mathrm{A}$                         & $\mathrm{B}$
  & $\mathrm{A}\oplus\mathrm{B}$
  \\
  $\mathbf{C}_{2(a)},\mathbf{C}_{2(b)}$
  & $\mathrm{A}$                         & $\mathrm{B}$
  & $\mathrm{B}$                         & $\mathrm{A}$
  & $\mathrm{A}\oplus\mathrm{B}$
  \\ \hline \hline
\end{tabular}
\label{T:CompatibilityRelationC4vC2vC2}
\end{table}

\clearpage